\documentclass[a4paper, 12pt]{article}

\pdfoutput = 1

\usepackage{styleBuding}
\usepackage{bm}
\usepackage{color,listings}
\usepackage[usenames,dvipsnames,svgnames,table]{xcolor}
\usepackage{amsthm, amsmath}
\usepackage{amssymb}
\usepackage{mathrsfs}
\usepackage{graphicx}
\usepackage{slashed}
\usepackage{soul}
\usepackage{multirow}
\usepackage{subfigure}
\usepackage{siunitx} 
\usepackage{url} 
\usepackage[utf8]{inputenc}
\usepackage[figuresright]{rotating}
\definecolor{back}{HTML}{F8F8F8}
\usepackage{epstopdf}

\newcommand{\rom}[1]{\uppercase\expandafter{\romannumeral #1\relax}}

\allowdisplaybreaks
\DeclareUnicodeCharacter{2212}{-}

\let\jnfont=\rm
\def\NPB#1,{{\jnfont Nucl.\ Phys.\ B }{\bf #1},}
\def\PLB#1,{{\jnfont Phys.\ Lett.\ B }{\bf #1},}
\def\EPJC#1,{{\jnfont Eur.\ Phys.\ Jour.\ C }{\bf #1},}
\def\PRD#1,{{\jnfont Phys.\ Rev.\ D }{\bf #1},}
\def\PRL#1,{{\jnfont Phys.\ Rev.\ Lett.\ }{\bf #1},}
\def\MPLA#1,{{\jnfont Mod.\ Phys.\ Lett.\ A }{\bf #1},}
\def\JPG#1,{{\jnfont J.\ Phys.\ G}{\bf #1},}
\def\CTP#1,{{\jnfont Commun.\ Theor.\ Phys.\ }{\bf #1},}
\def\ZPC#1,{{\jnfont Z.\ Phys.\ C }{\bf #1},}
\def\JHEP#1,{{\jnfont JHEP \ }{\bf #1},}

\title{\boldmath Electron and Muon Anomalous  Magnetic Moment in the $\mathbb{Z}_3$-NMSSM}

\author{Junjie Cao$^{a,b}$, Lei Meng$^a$, Yuanfang Yue$^{a}$}

\affiliation{ $^a$ Department of Physics, Henan Normal University, Xinxiang 453007, China}
\affiliation{ $^b$ Schools of Physics, Shandong University, Jinan, Shandong 250100, China}
\emailAdd{junjiec@alumni.itp.ac.cn}
\emailAdd{mel18@foxmail.com}
\emailAdd{yueyuanfang@htu.edu.cn}

\abstract{Inspired by the recent measurements of the muon and electron anomalous magnetic moments, the rapid progress of the LHC search for supersymmetry, and the significantly improved sensitivities of dark matter direct detection experiments, we studied the supersymmetric contribution to the electron \texorpdfstring{$g-2$}{}, $a_e^{\rm SUSY}$, in the Next-to-Minimal Supersymmetric Standard Model with a discrete $\mathbb{Z}_3$ symmetry. We concluded that $a_e^{\rm SUSY}$ was mainly correlated with $a_\mu^{\rm SUSY}$ by the formula $a_e^{\rm SUSY}/m_e^2 \simeq a_\mu^{\rm SUSY}/m_\mu^2$, and significant violations of this correlation might occur only in rare cases. As a result, $a_e^{\rm SUSY}$ was typically around $5 \times 10^{-14}$ when $a_\mu^{\rm SUSY} \simeq 2.5 \times 10^{-9}$. We also concluded that the dark matter direct detection and LHC experiments played crucial roles in determining the maximum reach of $a_e^{\rm SUSY}$. Concretely, $a_e^{\rm SUSY}$ might be around $3 \times 10^{-13}$ in the optimum cases if one used the XENON-1T experiment to limit the supersymmetry parameter space. This prediction, however, was reduced to $1.5 \times 10^{-13}$ after implementing the LZ restrictions and $1.0 \times 10^{-13}$ when further considering the LHC restrictions. }

\begin{document}
    \maketitle
    \flushbottom
\newpage
\section{Introduction}
The anomalous magnetic moment of leptons, $a_{l} \equiv g_{l}-2$, is one of the most accurate measurements in particle physics. In 2021, the E989 experiment at FermiLab published its first result of the muon anomalous magnetic moment~\cite{Abi:2021gix}, which consistent with the previous measurement of the Brookhaven National Laboratory (BNL) E821 experiment~\cite{Bennett:2006fi,Aoyama:2020ynm}. The combined result revealed a $4.2 \sigma $ deviation from the prediction of the Standard Model (SM)~\cite{Aoyama:2012wk,Aoyama:2019ryr,Czarnecki:2002nt,Gnendiger:2013pva,Davier:2017zfy,Keshavarzi:2018mgv,Colangelo:2018mtw,Hoferichter:2019gzf,Davier:2019can,Keshavarzi:2019abf,Kurz:2014wya,Melnikov:2003xd,Masjuan:2017tvw,Colangelo:2017fiz,Hoferichter:2018kwz,Gerardin:2019vio,Bijnens:2019ghy,Colangelo:2019uex,Blum:2019ugy,Colangelo:2014qya}:
\begin{eqnarray}\label{equ:deltaamu}
\Delta a_{\mu} \equiv a^{\rm Exp}_\mu - a^{\rm SM}_{\mu} = (25.1 \pm 5.9) \times 10^{-10},
\end{eqnarray}
if the recent lattice calculations of the hadron vacuum polarization (HVP) were ignored~\cite{Borsanyi:2020mff,Alexandrou:2022amy,Ce:2022kxy}
\footnote{Roughly speaking, these calculations can mitigate the discrepancy~\cite{Aoyama:2020ynm}. However, they are in tensions with the HVP contribution extracted from the $e^+ e^-$ data and the global fit of the precision electroweak observations~\cite{Crivellin:2020zul,Colangelo:2022vok,deRafael:2020uif,Keshavarzi:2020bfy}.}. Remarkably, the uncertainty in  $\Delta a_{\mu}$ will significantly reduce in the coming times when more experimental data from ongoing experiments at Fermilab~\cite{Muong-2:2021vma} and future experiments at JPARC~\cite{Abe:2019thb} are available. As a result, a more than $5\sigma$ deviation may be achieved if the central values of $a_\mu^{\rm SM}$ and $a_\mu^{\rm Exp}$ remain
unchanged, which will serve as a robust evidence of new physics beyond the SM (BSM). In addition, the measurement of the fine structure constant by the Laboratoire Kastler Brossel (LKB) using $^{87}$Rb atoms~\cite{Morel:2020dww} concluded a positive difference of $1.6\sigma$ between the experimental measurement and the SM prediction of electron anomalous magnetic moment~\cite{Hanneke:2008tm}:
\begin{eqnarray}\label{equ:deltaae1}
\Delta a_e^{\rm Rb} \equiv a^{\rm Exp, Rb}_e - a^{\rm SM}_e = (4.8 \pm 3.0) \times 10^{-13},
\end{eqnarray}
while that by the Lawrence Berkeley National Laboratory (LBNL) using the $^{133}$Cs atom obtained a negative difference of $2.4 \sigma$~\cite{2018Sci...360..191P}:
\begin{eqnarray}\label{equ:deltaae2}
\Delta a_e^{\rm Cs} \equiv a^{\rm Exp, Cs}_e - a^{\rm SM}_e = (-8.7 \pm 3.6) \times 10^{-13}.
\end{eqnarray}
Although the two results differ by more than $5 \sigma$ for still unknown reasons~\cite{Morel:2020dww}, they all indicate a sizable deviation from $a_e^{\rm SM}$ and thus, like the muon \texorpdfstring{$g-2$}{} anomaly, hint at the BSM physics' existence.

To date, the muon \texorpdfstring{$g-2$}{} anomaly has been intensively studied, which was recently reviewed in Ref.~\cite{Aoyama:2020ynm,Athron:2021iuf}. In contrast, only a few unified explanations of the electron and muon anomalies were investigated in the extensions of the SM with $SU(2)_L$ singlet or doublet Higgs bosons~\cite{Davoudiasl:2018fbb,Liu:2018xkx,Bauer:2019gfk,Cornella:2019uxs,Endo:2020mev,Haba:2020gkr,Adhikari:2021yvx,Bauer:2021mvw,De:2021crr,Chua:2020dya,Gardner:2019mcl,Dutta:2020scq,Botella:2020xzf,Jana:2020pxx,Han:2018znu,Hernandez:2021tii,DelleRose:2020oaa,Li:2020dbg,Han:2021gfu,Barman:2021xeq,Botella:2022eyw,Hue:2021xzl,Fajfer:2021cxa},
vector-like fermions~\cite{Crivellin:2018qmi,Chun:2020uzw,Chen:2020tfr,Hati:2020fzp,Jana:2020joi,Hiller:2019mou,Borah:2021khc,Biswas:2021dan,Botella:2022rte,Bharadwaj:2021tgp,Crivellin:2019mvj},
Leptoquarks~\cite{Keung:2021rps,Bigaran:2020jil,Dorsner:2020aaz,Bigaran:2021kmn}, different seesaw mechanisms~\cite{Escribano:2021css,Chowdhury:2022jde,Hong:2022xjg,Hernandez:2021xet,Mondal:2021vou,Arbelaez:2020rbq}, and new gauge symmetries~\cite{Abdullah:2019ofw,CarcamoHernandez:2020pxw,CarcamoHernandez:2019ydc,Bodas:2021fsy,Chowdhury:2021tnm,Hernandez:2021iss,Chen:2020jvl,Calibbi:2020emz,Anastasopoulos:2022ywj},
respectively, and their supersymmetric versions~\cite{Yang:2020bmh,Badziak:2019gaf,Endo:2019bcj,Dutta:2018fge,Banerjee:2020zvi,Frank:2021nkq,Li:2021koa,Li:2021xmw,Ali:2021kxa,Li:2022zap,Dao:2022rui,Giudice:2012ms,Dong:2019iaf,Cao:2021lmj}. The anomalies were also studied in the effective lagrangian framework~\cite{Aebischer:2021uvt}. These researches revealed that explaining the two anomalies together in the same new physical model was challenging if there were no flavor mixings in the lepton part. This is because the BSM contribution to the lepton anomalous magnetic moment can be decomposed as the square of the lepton mass multiplied by the factor $R_\ell$ in the flavor conserving case. Eqs.~\ref{equ:deltaamu},~\ref{equ:deltaae1} and ~\ref{equ:deltaae2} then indicate that $R_\ell$ for electron and muon are
\begin{eqnarray}
R_e^{\rm Rb} \equiv \frac{\Delta a_e^{\rm Rb}}{m_e^2} &=& 1.85\times 10^{-6}~{\rm GeV}^{-2}, \label{equ:Rme1} \\
R_e^{\rm Cs} \equiv \frac{\Delta a_e^{\rm Cs}}{m_e^2} &=& -3.34 \times 10^{-6}~{\rm GeV}^{-2}, \label{equ:Rme2} \\
R_\mu \equiv \frac{\Delta a_\mu}{m_\mu^2} &=& 2.25\times 10^{-7}~{\rm GeV}^{-2},  \label{equ:Rmu}
\end{eqnarray}
respectively. Since $|R_e|$ is at least 7 times larger than $R_\mu$, it is difficult to explain the two anomalies by a common physical origin. In the following, we concentrate on $\Delta a_e^{\rm Rb}$ instead of $\Delta a_e^{\rm Cs}$ in seeking for the common origin because it is easier to be accounted for by the BSM effects.

Among the new physics theories that can provide a consistent description of the leptonic anomalies, the minimal realizations of supersymmetry (SUSY) are most attractive due to their elegant theoretical structure and capabilities of naturally solving many problems of the SM, such as the hierarchy problem, the unification of the three forces, and the mystery of dark matter (DM)~\cite{Fayet:1976cr,Haber:1984rc,Gunion:1984yn,Djouadi:2005gj,Martin:1997ns,Jungman:1995df}. Studies of the muon \texorpdfstring{$g-2$}{} anomaly have revealed that $\Delta a_\mu$ can be totally contributed to by smuon-neutralino and sneutrino-chargino loops~\cite{Martin:2001st,Domingo:2008bb,Moroi:1995yh,Hollik:1997vb,Athron:2015rva,Endo:2021zal,Stockinger:2006zn,Czarnecki:2001pv,Cao:2011sn,Kang:2016iok,Zhu:2016ncq,Yanagida:2017dao, Hagiwara:2017lse,Cox:2018qyi,Tran:2018kxv,Padley:2015uma,Choudhury:2017fuu,Okada:2016wlm,Du:2017str, Ning:2017dng, Wang:2018vxp,Yang:2018guw,Liu:2020nsm,Cao:2019evo,Cao:2021lmj,Ke:2021kgy,Lamborn:2021snt,Li:2021xmw,Nakai:2021mha,Li:2021koa,Kim:2021suj,Li:2021pnt,Altmannshofer:2021hfu,Baer:2021aax,Chakraborti:2021bmv,Aboubrahim:2021xfi,Iwamoto:2021aaf,Chakraborti:2021dli,Cao:2021tuh,Yin:2021mls,Zhang:2021gun,Ibe:2021cvf,Han:2021ify,Wang:2021bcx,Zheng:2021gug,Chakraborti:2021mbr,Aboubrahim:2021myl,Ali:2021kxa,Wang:2021lwi,Chakraborti:2020vjp,Baum:2021qzx,Cao:2022chy,Cao:2022htd,Domingo:2022pde,Cao:2022ovk,He:2023lgi}.
Since these contributions contain a chiral enhancement factor $\tan \beta$, the involved sleptons and electroweakinos may be sufficiently heavy to coincide with the LHC search for SUSY~\cite{He:2023lgi}. Joint interpretations of both anomalies were recently discussed in the Minimal Supersymmetric Standard Model (MSSM)~\cite{Badziak:2019gaf,Dutta:2018fge,Li:2021koa,Li:2021xmw,Li:2022zap}. It was found that even without the flavor violation, the theory could explain the anomalies at $1 \sigma$ level by adjusting the magnitudes of the bino-slepton and chargino-sneutrino contributions for the electron and muon sectors~\cite{Badziak:2019gaf}. The favored parameter space was characterized by the selectrons and wino-like electroweakinos lighter than about $200~{\rm GeV}$ and the higgsinos heavier than about $1~{\rm TeV}$. These regions, however, have been excluded by the LHC search for SUSY according to our detailed Monte Carlo simulations~\cite{He:2023lgi}. Furthermore, given the general flavor structure of SUSY-breaking terms, the contribution of large non-universal trilinear A-terms could give the correct effect in principle~\cite{Crivellin:2018qmi,Dutta:2018fge,Giudice:2012ms}. Unfortunately, it seemed challenging to implement such anarchic A-terms with the SUSY-breaking mechanism while respecting all other flavor constraints~\cite{Crivellin:2018qmi}. These conclusions motivate us to conduct a more comprehensive study of the anomalies in the minimal realizations and explore as many possibilities of the theories as possible.

Notably, the direct detections of DM by the LUX-ZEPLIN (LZ) experiment~\cite{LZ:2022ufs} and the LHC search for SUSY~\cite{ATLAS:2019lff,ATLAS:2021moa,CMS:2020bfa,CMS:2018szt,CMS:2017moi,ATLAS:2018ojr,ATLAS:2018eui,ATLAS:2020pgy,ATLAS:2018qmw,CMS:2017kxn,CMS:2017bki,CMS:2018xqw,ATLAS:2018nud,ATLAS:2021yyr,ATLAS:2019lng,ATLAS:2017vat,CMS:2018kag,CMS:2018eqb} have required the higgsino mass in the MSSM to be significantly higher than the electroweak scale, namely, $\mu \gtrsim 500~{\rm GeV}$, in explaining the muon \texorpdfstring{$g-2$}{} anomaly at the $2\sigma$ level~\cite{He:2023lgi}. Although such a large $\mu$ may be generated  by the well-known Giudice-Masiero mechanism in the gravity-mediated SUSY breaking scenario~\cite{Giudice:1988yz}, it usually induces severe fine-tuning problems in the light of the LHC Higgs discovery and the absence of any SUSY discovery when the theory runs down from an infrared high energy scale to the electroweak scale~\cite{Arvanitaki:2013yja,Evans:2013jna,Baer:2014ica}. Given the unnaturalness of the MSSM, we focus on the low energy Next-to-Minimal Supersymmetric Standard Model with a discrete $\mathbb{Z}_3$ symmetry ($\mathbb{Z}_3$-NMSSM), which is another economic realization of SUSY~\cite{Ellwanger:2009dp,Maniatis:2009re}, to study the two anomalies. This model extends the MSSM with a singlet superfield $\hat{S}$ to dynamically generate the $\mu$-parameter of the MSSM after the scalar component field of $\hat{S}$ develops a vacuum expectation value (vev) of ${\cal{O}}(1~{\rm TeV})$. Consequently, this theory is self-contained at the electroweak scale and owns much richer phenomenology than the MSSM as indicated by, e.g., Ref.~\cite{Ellwanger:2011aa,Cao:2012fz,Ellwanger:2014hia,Cao:2018rix,Baum:2017enm}. As a preliminary study on this subject in the $\mathbb{Z}_3$-NMSSM, we assume no flavor violation in the lepton sector to simplify this research.

In our previous work~\cite{Cao:2022htd}, we investigated the impacts of the muon \texorpdfstring{$g-2$}{} anomaly on the $\mathbb{Z}_3$-NMSSM by including the restrictions from the DM relic density~\cite{Planck:2018vyg}, the direct detection of DM by the XENON-1T experiments~\cite{Aprile:2018dbl,Aprile:2019dbj}, and the LHC search for SUSY~\cite{ATLAS:2019lff,ATLAS:2021moa,CMS:2020bfa,CMS:2018szt,CMS:2017moi,ATLAS:2018ojr,ATLAS:2018eui,ATLAS:2020pgy,ATLAS:2018qmw,CMS:2017kxn,CMS:2017bki,CMS:2018xqw,ATLAS:2018nud,ATLAS:2021yyr,ATLAS:2019lng,ATLAS:2017vat,CMS:2018kag,CMS:2018eqb} . We found that neither overly heavy supersymmetric particles nor moderately light sparticles were favored to explain the muon anomaly. In particular, we first obtained lower bounds on some SUSY parameters from those experimental restrictions, e.g., $|M_1| \gtrsim 275~{\rm GeV}$, $M_2 \gtrsim 300~{\rm GeV}$,  $\mu \gtrsim 460~{\rm GeV}$, $m_{\tilde{\mu}_L} \gtrsim 310~{\rm GeV}$, and $m_{\tilde{\mu}_R} \gtrsim 350~{\rm GeV}$, where $M_1$ and $M_2$ denoted the gaugino masses defined at the renormalization scale of $1~{\rm TeV}$ and $m_{\tilde{\mu}_L}$ and $m_{\tilde{\mu}_R}$ were the masses of the muon-type sletpons with L and R denoting their dominant chiral components, respectively. We also concluded by calculating the Bayesian evidence that the preferred DM candidate
was the bino-dominated lightest neutralino rather than the singlino-dominated neutralino. It mainly co-annihilated with the wino-dominated electroweakinos or the muon-type sleptons  to obtain the measured density. In this work, we improve the previous study by freely varying the masses of the electron-type sleptons to predict a sizable $a_e^{\rm SUSY}$, which represents the SUSY contribution to $a_e$, and subsequently restricting the theory with the latest experimental results. We find that the LZ experiment and the LHC search for SUSY are complementary to each other in limiting the parameter space. The restrictions are so tight that they reduce $a_e^{\rm SUSY}$ from $3 \times 10^{-13}$ to at most $1 \times 10^{-13}$ when $a_\mu^{\rm SUSY}$ is fixed at $2.5 \times 10^{-9}$. We also find some subtleties about the results of the previous study after comparing them with those of this research.

This work is organized as follows. In Sec.~\ref{sec:model}, we briefly recapitulated the dominant contributions to the lepton \texorpdfstring{$g-2$}{} in the $\mathbb{Z}_3$-NMSSM and the status of the LHC search for SUSY.
We studied the impacts of the leptonic anomalies on the $\mathbb{Z}_3$-NMSSM in Sec.~\ref{sec:na} and compared them with those of Ref.~\cite{Cao:2022htd}, where we only considered the muon \texorpdfstring{$g-2$}{} anomaly. After including relevant experimental constraints, we also presented the theory's electron and muon \texorpdfstring{$g-2$}{} prediction. We summarized the conclusions in Sec.~\ref{sec:sum}.

\section{\label{sec:model}Theoretical Preliminaries }

\subsection{Basics of the $\mathbb{Z}_3$-NMSSM}

The $\mathbb{Z}_3$-NMSSM contains one extra singlet Higgs superfield $\hat{S}$ besides the usual two Higgs doublets, $\hat{H}_u$ and $\hat{H}_d$, of the MSSM. Consequently, the associated superpotential and soft SUSY breaking lagrangian of the $\mathbb{Z}_3$-NMSSM are given by~\cite{Ellwanger:2009dp,Maniatis:2009re}
\begin{eqnarray}
W_{\rm NMSSM} &=& W_{\rm MSSM} + \lambda \hat{S} \hat{H}_u \cdot \hat{H}_d + \frac{1}{3}\kappa \hat{S}^3,\\ \nonumber
-\mathcal{L}_{\rm soft} &=&  \left[ A_{\lambda} \lambda  S H_u \cdot H_d + \frac{1}{3}  A_{\kappa} \kappa S^3 + h.c. \right]\\
&~&+m_{H_u}^2 |H_u|^2 + m_{H_d}^2 |H_d|^2 +m_{S}^2 |S|^2 +\cdots.
\end{eqnarray}
The $W_{\rm MSSM}$ is the MSSM superpotential without the $\mu$-term, $\lambda$ and $\kappa$ are dimensionless coefficients that parameterize the strength of Higgs self couplings, $H_u$, $H_d$, and $S$ are the scalar parts of the superfields $\hat H_u$, $\hat H_d$, and $\hat S$, respectively, and the dimensional quantities $m_{H_u}^2$, $m_{H_d}^2$, $m_{S}^2$, $A_{\lambda}$, and $A_{\kappa}$ are soft-breaking parameters. After the electroweak symmetry breaking, the fields $H_u$, $H_d$, and $S$ acquire the vevs $v_u/\sqrt{2}$, $v_d/\sqrt{2}$, and $v_s/\sqrt{2}$, respectively, with $v \equiv \sqrt{v_u^2 + v_d^2} = 246~\rm GeV$, and the interaction $\lambda \hat S \hat H_u \cdot \hat H_d $ generates an effective $\mu$-term with  $\mu = \lambda v_s/\sqrt{2}$. If the three vevs replace $m_{H_u}^2$, $m_{H_d}^2$, $m_{S}^2$ as theoretical inputs by the minimization conditions of the Higgs potential, the free parameters in the Higgs sector can be taken as follows~\cite{Ellwanger:2009dp}:
\begin{eqnarray}
\lambda,~~~\kappa,~~~A_\lambda,~~~A_{\kappa},~~~\mu ,~~~\rm \tan\beta \equiv v_u/v_d.
\end{eqnarray}

In the field convention that $H_{\rm SM} \equiv \sin \beta \rm Re(H^0_u )+\cos \beta \rm Re(H^0_d )$,  $H_{\rm NSM} \equiv \cos \beta \rm Re(H^0_u )- \sin \beta \rm Re(H^0_d )$, and $A_{\rm NSM} \equiv \cos \beta \rm Im(H^0_u ) - \sin\beta \rm Im(H^0_d )$,  the elements of the CP-even Higgs boson mass matrix $\mathcal{M}^2_S$ in the bases ($H_{\rm NSM}$, $H_{\rm SM}$, $\rm Re(S)$) are read as
\begin{eqnarray}
M_{S,11}^2 &=&  \frac{2\mu (A_\lambda +\kappa v_s)}{\sin 2\beta} + \frac{1}{2}(2m^2_Z -\lambda^2 v^2) \sin^2 2\beta, \quad M_{S,12}^2 =  -\frac{1}{4}(2m^2_Z-\lambda^2 v^2)\sin4\beta, \nonumber \\
M_{S,13}^2 &=&  -\frac{1}{\sqrt{2}}(\lambda A_{\lambda}+ 2 \kappa \mu)  v\cos2\beta, \quad M_{S,22}^2 =  m_Z^2\cos^2 2\beta +\frac{1}{2} \lambda^2v^2\sin^2 2\beta, \nonumber \\
M_{S,23}^2 &=& \frac{v}{\sqrt{2}}\left[ 4\lambda \mu-(\lambda A_\lambda+2\kappa\mu)\sin2\beta \right], \\
M_{S,33}^2 &=&  \frac{\lambda A_\lambda \sin 2\beta}{4\mu}\lambda v^2  +\kappa A_{\kappa} v_s +4(\kappa v_s)^2 - \frac{1}{2}\lambda^2 v^2. \nonumber
\end{eqnarray}
Then the mixings of the fields $H_{\rm NSM}$, $H_{\rm SM}$, and ${\rm Re}(S)$ result in three CP-even mass eigenstates denoted by $h_i$  with $i = 1, 2, 3$ and satisfying $m_{h_1} < m_{h_2} < m_{h_3}$. Similarly, the mixing of $A_{\rm NSM}$ and ${\rm Im}(S)$ leads to two CP-odd states $A_j$ with $j=1, 2$ and satisfying $m_{A_1}<m_{A_2}$. The model also predicts a pair of charged Higgs bosons $H^{\pm} \equiv \cos \beta H^{\pm}_ u + \sin \beta H^{\pm}_d$. In this study, the lightest CP-even Higgs boson $h_1$, instead of the next-lightest Higgs boson $h_2$, is often treated as the LHC-discovered Higgs boson, denoted as $h$ in this research, since the Bayesian evidence of the $h_1$-scenario is much
larger than that of the $h_2$-scenario after considering the Higgs data collected at the LHC~\cite{Cao:2022htd}.

The electroweakino sector comprises bino ($\tilde{B}$), wino ($\tilde{W}$), higgsino ($\tilde{H}_u$ and $\tilde{H}_d$), and singlino ($\tilde{S}$) fields. Their mixings lead to five neutralinos $\tilde{\chi}^0_i$ with $i=1,2,...,5$, arranged in ascending mass order, and two charginos  $\tilde{\chi}^\pm_j$ with $j=1,2$ and satisfying $m_{\tilde{\chi}^\pm_1}<m_{\tilde{\chi}^\pm_2}$~\cite{Ellwanger:2009dp}. Their masses and mixings are determined by the parameters $M_1$, $M_2$, $\lambda$, $\kappa$, $\tan \beta$, and $\mu$, where the last four parameters also appear in the Higgs mass matrices.

\subsection{\label{sec:22}Leptonic \texorpdfstring{$g-2$}{} in the $\mathbb{Z}_3$-NMSSM}

The SUSY effects on the lepton anomalous magnetic moment $a_\ell$ ($\ell = e, \mu$) arise from the loops containing a chargino and a sneutrino and those mediated by a neutralino and a slepton~\cite{Domingo:2008bb,Martin:2001st}. The one-loop expressions of $a_\ell$ are~\cite{Domingo:2008bb}
\begin{eqnarray}\begin{split}
a_{\ell}^{\rm SUSY} &= a_{\ell}^{\tilde{\chi}^0 \tilde{\ell}} + a_{\ell}^{\tilde{\chi}^{\pm} \tilde{\nu}},\\
a_{\ell}^{\tilde{\chi}^0 \tilde{\ell}} &= \frac{m_{\ell}}{16 \pi^2}\sum_{i,k}\left\{
-\frac{m_{\ell}}{12 m_{\tilde{\ell}_k}^2} \left( |n_{ik}^{\rm L}|^2 + |n_{ik}^{\rm R}|^2 \right) F_1^{\rm N}(x_{ik}) + \frac{m_{\tilde{\chi}_i^0}}{3 m_{\tilde{\ell}_k}^2} {\rm Re}(n_{ik}^{\rm L} n_{ik}^{\rm R}) F_2^{\rm N}(x_{ik})
\right\}, \nonumber \\
a_{\ell}^{\tilde{\chi}^\pm \tilde{\nu}} &= \frac{m_{\ell}}{16 \pi^2}\sum_{j}\left\{
\frac{m_{\ell}}{12 m_{\tilde{\nu}_{\ell}}^2} \left( |c_{j}^{\rm L}|^2 + |c_{j}^{\rm R}|^2 \right) F_1^{\rm C}(x_{j}) + \frac{2 m_{\tilde{\chi}_j^\pm}}{3 m_{\tilde{\nu}_{\ell}}^2} {\rm Re}(c_{j}^{\rm L}c_{j}^{ \rm R}) F_2^{\rm C}(x_{j})
\right\},
\end{split}
\end{eqnarray}
where  $i=1,...,5$, $j=1,2$, and $k=1,2$ refer to the neutralino, chargino, and slepton indices, respectively. The involved couplings are given by
\begin{equation}
\begin{split}
n_{ik}^{\rm L} 	= \frac{1}{\sqrt{2}}\left( g_2 N_{i2} + g_1 N_{i1} \right)X^{\ell,*}_{k1} -Y_{\ell} N_{i3}X^{\ell,*}_{k2}, & \quad c_{j}^{\rm L} = -g_2 V_{j1}, \\
n_{ik}^{\rm R} = \sqrt{2} g_1 N_{i1} X_{k2}^\ell + Y_{\ell} N_{i3} X_{k1}^\ell, \quad
&c_{j}^{\rm R} = Y_{\ell} U_{j2}, \\
\end{split}
\end{equation}
where $N$, $X$, and $U$ and $V$ are the rotation matrices of the neutralinos, the sleptons, and the charginos, respectively~\cite{Ellwanger:2009dp}. The kinematic loop functions $F_i (x)$ take the following forms:
\begin{equation}
\begin{split}
F_1^{\rm N}(x) &= \frac{2}{(1-x)^4} \left( 1 - 6x + 3x^2 + 2x^3 - 6x^2 \ln{x} \right),\\
F_2^{\rm N}(x) &= \frac{3}{(1-x)^3} \left( 1 - x^2 + 2x \ln{x} \right),\\
F_1^{\rm C}(x) &= \frac{2}{(1-x)^4} \left( 2 + 3x - 6x^2 + x^3 + 6x \ln{x} \right), \\
F_2^{\rm C}(x) &= -\frac{3}{2(1-x)^3} \left( 3 - 4x + x^2 + 2\ln{x} \right),
\end{split}
\end{equation}
with $x_{ik} \equiv m_{\tilde{\chi}^0_i}^2 / m_{\tilde{\ell}_k}^2$ and $x_{j} \equiv m_{\tilde{\chi}^\pm_j}^2 / m_{\tilde{\nu}_\ell}^2$, and they satisfy $F_i (1) = 1$.

It is instructive to understand the behavior of $a_\ell^{\rm SUSY}$ by the mass insertion approximation~\cite{Moroi:1995yh}. In this method, the SUSY contributions to $a_\ell$ are classified into four types: "WHL", "BHL", "BHR", and "BLR", where $W$, $B$, $H$, $L$, and $R$ stand for wino, bino, higgsino, left-handed slepton or sneutrino, and right-handed slepton fields, respectively.  They arise from the Feynman diagrams involving $\tilde{W}-\tilde{H}_d$, $\tilde{B}-\tilde{H}_d^0$, $\tilde{B}-\tilde{H}_d^0$, and $\tilde{\ell}_L-\tilde{\ell}_R$ transitions, respectively, and have the following form~\cite{Athron:2015rva, Moroi:1995yh,Endo:2021zal}:
\begin{eqnarray}
a_{\ell, \rm WHL}^{\rm SUSY}
&=&\frac{\alpha_2}{8 \pi} \frac{m_{\ell}^2 M_2 \mu \tan \beta}{M_{\tilde{\nu}_\ell}^4} \left \{ 2 f_C\left(\frac{M_2^2}{M_{\tilde{\nu}_{\ell}}^2}, \frac{\mu^2}{M_{\tilde{\nu}_{\ell}}^2} \right) - \frac{M_{\tilde{\nu}_\ell}^4}{M_{\tilde{\ell}_L}^4} f_N\left(\frac{M_2^2}{M_{\tilde{\ell}_L}^2}, \frac{\mu^2}{M_{\tilde{\ell}_L}^2} \right) \right \}\,, \quad \quad
\\
a_{\ell, \rm BHL}^{\rm SUSY}
&=& \frac{\alpha_Y}{8 \pi} \frac{m_\ell^2 M_1 \mu  \tan \beta}{M_{\tilde{\ell}_L}^4} f_N\left(\frac{M_1^2}{M_{\tilde{\ell}_L}^2}, \frac{\mu^2}{M_{\tilde{\ell}_L}^2} \right)\,,
\\
a_{\ell, \rm BHR}^{\rm SUSY}
&=& - \frac{\alpha_Y}{4\pi} \frac{m_{\ell}^2 M_1 \mu \tan \beta}{M_{\tilde{\ell}_R}^4} f_N\left(\frac{M_1^2}{M_{\tilde{\ell}_R}^2}, \frac{\mu^2}{M_{\tilde{\ell}_R}^2} \right)\,,
\\
a_{\ell, \rm BLR}^{\rm SUSY}
&=& \frac{\alpha_Y}{4\pi} \frac{m_{\ell}^2  M_1 \mu \tan \beta}{M_1^4}
f_N\left(\frac{M_{\tilde{\ell}_L}^2}{M_1^2}, \frac{M_{\tilde{\ell}_R}^2}{M_1^2} \right)\,,
\end{eqnarray}
where $M_{\tilde{\ell}_L}$ and $M_{\tilde{\ell}_R}$ are soft-breaking masses for the left-handed and right-handed slepton fields, respectively, at the slepton mass scale, and they are approximately equal to the slepton masses.  The loop functions are given by
\begin{eqnarray}
f_C(x,y)
&=&  \frac{5-3(x+y)+xy}{(x-1)^2(y-1)^2} - \frac{2\ln x}{(x-y)(x-1)^3}+\frac{2\ln y}{(x-y)(y-1)^3} \,,
\\
f_N(x,y)
&=&
\frac{-3+x+y+xy}{(x-1)^2(y-1)^2} + \frac{2x\ln x}{(x-y)(x-1)^3}-\frac{2y\ln y}{(x-y)(y-1)^3} \,,
\end{eqnarray}
and they have the property that $f_C(1,1) = 1/2$ and $f_N(1,1) = 1/6$.

The following properties of $a_\ell^{\rm SUSY}$ should be noted:
\begin{itemize}
\item If all the dimensional SUSY parameters involved in $a_\ell^{\rm SUSY}$ take a common value $M_{\rm SUSY}$, $a_\ell^{\rm SUSY}$ is proportional to $m_\ell^2 \tan \beta/M_{\rm SUSY}^2$. In this case, $a_e^{\rm SUSY}$ and  $a_\mu^{\rm SUSY}$ are correlated by $a_e^{\rm SUSY}/a_\mu^{\rm SUSY} = m_e^2/m_\mu^2$, and $a_\mu^{\rm SUSY} = (25.1,~25.1-5.9,~ 25.1-2\times 5.9) \times 10^{-10}$ corresponds to $a_e^{\rm SUSY} = (5.85, 4.47, 3.10) \times 10^{-14}$, respectively. This conclusion reflects that the electron \texorpdfstring{$g-2$}{} anomaly prefers a much lower SUSY scale than the muon \texorpdfstring{$g-2$}{} anomaly.

\item The "WHL" contribution in each $a_\ell^{\rm SUSY}$ is usually much larger than the other contributions if $\tilde{\ell}_L$ is not significantly heavier than $\tilde{\ell}_R$~\cite{Cao:2021tuh}.

\item The four types of contributions have different dependence on the parameter $\mu$. Specifically, $a_{\ell, \rm BLR}^{\rm SUSY}$ is proportional to the higgsino mass $\mu$, while the others contain both a pre-factor of $\mu$ and a loop function that tends to be zero as $\mu$ approaches infinity. We observe that the "WHL" contribution monotonically decreases with increasing $\mu$ for several typical cases of particle mass spectra. By contrast, the "BHL" and "BHR" contributions increase when $\mu$ is significantly smaller than the slepton masses and decrease if $\mu$ is larger than the slepton masses.

\item Since the singlino field appears in the "WHL," "BHL", and "BHR" loops by two-time insertions, its contribution to $a_{\ell}^{\rm SUSY}$ is never prominent, considering $\lambda \lesssim 0.3$ for most cases in this study. Therefore, $a_{\ell}^{\rm SUSY}$ in the $\mathbb{Z}_3$-NMSSM is almost equal to that in the MSSM.

\item The difference of $a_\ell^{\rm SUSY}$ calculated by the mass insertion approximation and the full expression, respectively, is less than $3\%$. We verified this point by studying the samples acquired by the following parameter scan.

\item The two-loop (2L) contributions to $a_\ell^{\rm SUSY}$, including 2L corrections to the SM one-loop diagrams and those to the SUSY one-loop diagrams~\cite{Stockinger:2006zn}, are about $-5\%$ of the one-loop prediction~\cite{Cao:2022ovk}. They were neglected in this study.
\end{itemize}

\begin{table}[]
	\caption{Experimental analyses of the electroweakino production processes at the 13 TeV LHC, categorized by the topologies of the SUSY signals. They were utilized  to limit the parameter points of this research. }
	\label{Table1}
	\vspace{0.2cm}
	\resizebox{0.98\textwidth}{!}{
		\begin{tabular}{llll}
			\hline\hline
			\texttt{Scenario} & \texttt{Final State} &\multicolumn{1}{c}{\texttt{Name}}\\\hline
			\multirow{6}{*}{$\tilde{\chi}_{2}^0\tilde{\chi}_1^{\pm}\rightarrow WZ\tilde{\chi}_1^0\tilde{\chi}_1^0$}&\multirow{6}{*}{$n\ell (n\geq2) + nj(n\geq0) + \text{E}_\text{T}^{\text{miss}}$}&\texttt{CMS-SUS-20-001($137fb^{-1}$)}~\cite{CMS:2020bfa}\\&&\texttt{ATLAS-2106-01676($139fb^{-1}$)}~\cite{ATLAS:2021moa}\\&&\texttt{CMS-SUS-17-004($35.9fb^{-1}$)}~\cite{CMS:2018szt}\\&&\texttt{CMS-SUS-16-039($35.9fb^{-1}$)}~\cite{CMS:2017moi}\\&&\texttt{ATLAS-1803-02762($36.1fb^{-1}$)}~\cite{ATLAS:2018ojr}\\&&\texttt{ATLAS-1806-02293($36.1fb^{-1}$)}~\cite{ATLAS:2018eui}\\\\
			\multirow{2}{*}{$\tilde{\chi}_2^0\tilde{\chi}_1^{\pm}\rightarrow \ell\tilde{\nu}\ell\tilde{\ell}$}&\multirow{2}{*}{$n\ell (n=3) + \text{E}_\text{T}^{\text{miss}}$}&\texttt{CMS-SUS-16-039($35.9fb^{-1}$)}~\cite{CMS:2017moi}\\&&\texttt{ATLAS-1803-02762($36.1fb^{-1}$)}~\cite{ATLAS:2018ojr}\\\\
			$\tilde{\chi}_2^0\tilde{\chi}_1^{\pm}\rightarrow \tilde{\tau}\nu\ell\tilde{\ell}$&$2\ell + 1\tau + \text{E}_\text{T}^{\text{miss}}$&\texttt{CMS-SUS-16-039($35.9fb^{-1}$)}~\cite{CMS:2017moi}\\\\
			$\tilde{\chi}_2^0\tilde{\chi}_1^{\pm}\rightarrow \tilde{\tau}\nu\tilde{\tau}\tau$&$3\tau + \text{E}_\text{T}^{\text{miss}}$&\texttt{CMS-SUS-16-039($35.9fb^{-1}$)}~\cite{CMS:2017moi}\\\\
			\multirow{6}{*}{$\tilde{\chi}_{2}^0\tilde{\chi}_1^{\pm}\rightarrow Wh\tilde{\chi}_1^0\tilde{\chi}_1^0$}&\multirow{6}{*}{$n\ell(n\geq1) + nb(n\geq0) + nj(n\geq0) + \text{E}_\text{T}^{\text{miss}}$}&\texttt{ATLAS-1909-09226($139fb^{-1}$)}~\cite{ATLAS:2020pgy}\\&&\texttt{CMS-SUS-17-004($35.9fb^{-1}$)}~\cite{CMS:2018szt}\\&&\texttt{CMS-SUS-16-039($35.9fb^{-1}$)}~\cite{CMS:2017moi}\\
			&&\texttt{ATLAS-1812-09432($36.1fb^{-1}$)}\cite{ATLAS:2018qmw}\\&&\texttt{CMS-SUS-16-034($35.9fb^{-1}$)}\cite{CMS:2017kxn}\\&&\texttt{CMS-SUS-16-045($35.9fb^{-1}$)}~\cite{CMS:2017bki}\\\\
			\multirow{2}{*}{$\tilde{\chi}_1^{\mp}\tilde{\chi}_1^{\pm}\rightarrow WW\tilde{\chi}_1^0 \tilde{\chi}_1^0$}&\multirow{2}{*}{$2\ell + \text{E}_\text{T}^{\text{miss}}$}&\texttt{ATLAS-1908-08215($139fb^{-1}$)}~\cite{ATLAS:2019lff}\\&&\texttt{CMS-SUS-17-010($35.9fb^{-1}$)}~\cite{CMS:2018xqw}\\\\
			\multirow{2}{*}{$\tilde{\chi}_1^{\mp}\tilde{\chi}_1^{\pm}\rightarrow 2\tilde{\ell}\nu(\tilde{\nu}\ell)$}&\multirow{2}{*}{$2\ell + \text{E}_\text{T}^{\text{miss}}$}&\texttt{ATLAS-1908-08215($139fb^{-1}$)}~\cite{ATLAS:2019lff}\\&&\texttt{CMS-SUS-17-010($35.9fb^{-1}$)}~\cite{CMS:2018xqw}\\\\
			$\tilde{\chi}_2^{0}\tilde{\chi}_1^{\mp}\rightarrow h/ZW\tilde{\chi}_1^0\tilde{\chi}_1^0,\tilde{\chi}_1^0\rightarrow \gamma/Z\tilde{G}$&\multirow{2}{*}{$2\gamma + n\ell(n\geq0) + nb(n\geq0) + nj(n\geq0) + \text{E}_\text{T}^{\text{miss}}$}&\multirow{2}{*}{\texttt{ATLAS-1802-03158($36.1fb^{-1}$)}~\cite{ATLAS:2018nud}}\\$\tilde{\chi}_1^{\pm}\tilde{\chi}_1^{\mp}\rightarrow WW\tilde{\chi}_1^0\tilde{\chi}_1^0,\tilde{\chi}_1^0\rightarrow \gamma/Z\tilde{G}$&&\\\\
			$\tilde{\chi}_2^{0}\tilde{\chi}_1^{\pm}\rightarrow ZW\tilde{\chi}_1^0\tilde{\chi}_1^0,\tilde{\chi}_1^0\rightarrow h/Z\tilde{G}$&\multirow{4}{*}{$n\ell(n\geq4) + \text{E}_\text{T}^{\text{miss}}$}&\multirow{4}{*}{\texttt{ATLAS-2103-11684($139fb^{-1}$)}~\cite{ATLAS:2021yyr}}\\$\tilde{\chi}_1^{\pm}\tilde{\chi}_1^{\mp}\rightarrow WW\tilde{\chi}_1^0\tilde{\chi}_1^0,\tilde{\chi}_1^0\rightarrow h/Z\tilde{G}$&&\\$\tilde{\chi}_2^{0}\tilde{\chi}_1^{0}\rightarrow Z\tilde{\chi}_1^0\tilde{\chi}_1^0,\tilde{\chi}_1^0\rightarrow h/Z\tilde{G}$&&\\$\tilde{\chi}_1^{\mp}\tilde{\chi}_1^{0}\rightarrow W\tilde{\chi}_1^0\tilde{\chi}_1^0,\tilde{\chi}_1^0\rightarrow h/Z\tilde{G}$&&\\\\
			\multirow{3}{*}{$\tilde{\chi}_{i}^{0,\pm}\tilde{\chi}_{j}^{0,\mp}\rightarrow \tilde{\chi}_1^0\tilde{\chi}_1^0+\chi_{soft}\rightarrow ZZ/H\tilde{G}\tilde{G}$}&\multirow{3}{*}{$n\ell(n\geq2) + nb(n\geq0) + nj(n\geq0) + \text{E}_\text{T}^{\text{miss}}$}&\texttt{CMS-SUS-16-039($35.9fb^{-1}$)}~\cite{CMS:2017moi}\\&&\texttt{CMS-SUS-17-004($35.9fb^{-1}$)}~\cite{CMS:2018szt}\\&&\texttt{CMS-SUS-20-001($137fb^{-1}$)}~\cite{CMS:2020bfa}\\\\
			\multirow{2}{*}{$\tilde{\chi}_{i}^{0,\pm}\tilde{\chi}_{j}^{0,\mp}\rightarrow \tilde{\chi}_1^0\tilde{\chi}_1^0+\chi_{soft}\rightarrow HH\tilde{G}\tilde{G}$}&\multirow{2}{*}{$n\ell(n\geq2) + nb(n\geq0) + nj(n\geq0) + \text{E}_\text{T}^{\text{miss}}$}&\texttt{CMS-SUS-16-039($35.9fb^{-1}$)}~\cite{CMS:2017moi}\\&&\texttt{CMS-SUS-17-004($35.9fb^{-1}$)}~\cite{CMS:2018szt}\\\\
			$\tilde{\chi}_{2}^{0}\tilde{\chi}_{1}^{\pm}\rightarrow W^{*}Z^{*}\tilde{\chi}_1^0\tilde{\chi}_1^0$&$3\ell + \text{E}_\text{T}^{\text{miss}}$&\texttt{ATLAS-2106-01676($139fb^{-1}$)}~\cite{ATLAS:2021moa}\\\\
			\multirow{3}{*}{$\tilde{\chi}_{2}^{0}\tilde{\chi}_{1}^{\pm}\rightarrow Z^{*}W^{*}\tilde{\chi}_1^0\tilde{\chi}_1^0$}&\multirow{2}{*}{$2\ell + nj(n\geq0) + \text{E}_\text{T}^{\text{miss}}$}&\texttt{ATLAS-1911-12606($139fb^{-1}$)}~\cite{ATLAS:2019lng}\\&&\texttt{ATLAS-1712-08119($36.1fb^{-1}$)}~\cite{ATLAS:2017vat}\\&&\texttt{CMS-SUS-16-048($35.9fb^{-1}$)}~\cite{CMS:2018kag}\\\\
			\multirow{3}{*}{$\tilde{\chi}_{2}^{0}\tilde{\chi}_{1}^{\pm}+\tilde{\chi}_{1}^{\pm}\tilde{\chi}_{1}^{\mp}+\tilde{\chi}_{1}^{\pm}\tilde{\chi}_{1}^{0}$}&\multirow{3}{*}{$2\ell + nj(n\geq0) + \text{E}_\text{T}^{\text{miss}}$}&\texttt{ATLAS-1911-12606($139fb^{-1}$)}~\cite{ATLAS:2019lng}\\&&\texttt{ATLAS-1712-08119($36.1fb^{-1}$)}~\cite{ATLAS:2017vat}\\&&\texttt{CMS-SUS-16-048($35.9fb^{-1}$)}~\cite{CMS:2018kag}\\\hline
					
	\end{tabular}} 
\end{table}

\begin{table}[]
	\caption{Same as Table~\ref{Table1}, but for the slepton production processes.}
	\label{Table2}
  \centering
	\vspace{0.2cm}
	\resizebox{0.7\textwidth}{!}{
		\begin{tabular}{llll}
			\hline\hline
			\texttt{Scenario} & \texttt{Final State} &\multicolumn{1}{c}{\texttt{Name}}\\\hline
\multirow{6}{*}{$\tilde{\ell}\tilde{\ell}\rightarrow \ell\ell\tilde{\chi}_1^0\tilde{\chi}_1^0$}&\multirow{6}{*}{$2\ell + \text{E}_\text{T}^{\text{miss}}$}&\multirow{1}{*}{\texttt{ATLAS-1911-12606($139fb^{-1}$)}~\cite{ATLAS:2019lng}}\\&&\multirow{1}{*}{\texttt{ATLAS-1712-08119($36.1fb^{-1}$)}~\cite{ATLAS:2017vat}}\\&&\multirow{1}{*}{\texttt{ATLAS-1908-08215($139fb^{-1}$)}~\cite{ATLAS:2019lff}}\\&&\multirow{1}{*}{\texttt{CMS-SUS-20-001($137fb^{-1}$)}~\cite{CMS:2020bfa}}\\&&\multirow{1}{*}{\texttt{ATLAS-1803-02762($36.1fb^{-1}$)}~\cite{ATLAS:2018ojr}}\\&&\multirow{1}{*}{\texttt{CMS-SUS-17-009($35.9fb^{-1}$)}~\cite{CMS:2018eqb}}\\\hline

\end{tabular}} 
\end{table}

\subsection{LHC Search for SUSY}\label{LHC:Ana}

To explain the electron and muon \texorpdfstring{$g-2$}{} anomalies in the $\mathbb{Z}_3$-NMSSM simultaneously , both the electroweakinos and the first two-generation sleptons must be moderately light. The LHC experiments have strongly limited such a situation by searching for the multi-lepton plus missing momentum signal. We present pertinent experimental analyses in Tables \ref{Table1} and \ref{Table2}. Notably, the following ones play a crucial role in restricting the situation:

\begin{itemize}
\item \texttt{CMS-SUS-20-001~\cite{CMS:2020bfa}}: Search for the SUSY signal containing two oppositely charged same-flavor leptons and missing transverse momentum. This analysis investigated not only the squark and gluino productions but also the electroweakino and slepton productions. The lepton arised from an on-shell or off-shell $Z$ boson in the decay chain or from the direct decay of the produced sleptons. The wino-dominated chargino and neutralino were explored up to $750~{\rm GeV}$ and $800~{\rm GeV}$, respectively, in mass by the electroweakino pair production processes, while the first two-generation sleptons were explored up to a mass of $700~{\rm GeV}$ by the slepton pair production processes, assuming the sleptons were degenerate in mass.

\item \texttt{CMS-SUS-16-039 and CMS-SUS-17-004~\cite{CMS:2017moi,CMS:2018szt}}: Search for the electro-weakino productions by the final state containing two, three, or four leptons and missing transverse momentum ($\rm{E}_{\rm{T}}^{\rm{miss}}$). The analyses included all possible final states and defined several categories by the number of leptons in the event, their flavor and charges to improve the discovery potential. In the context of the simplified model of SUSY, the observed limit on the wino-dominated $m_{\tilde{\chi}_1^{\pm}}$ in the chargino-neutralino production was 650 GeV for the $WZ$ topology, 480 GeV for the $WH$ topology, and 535 GeV for the mixed topology. Remarkably, these analyses studied only $35.9~{\rm fb^{-1}}$ data collected at the Run-II phase of the LHC.

\item \texttt{ATLAS-2106-01676~\cite{ATLAS:2021moa}}: Search for the signals of the wino- and higgsino-dominated chargino-neutralino associated productions. It investigated on-shell $WZ$, off-shell $WZ$, and $Wh$ categories in the decay chain and concentrated on the final state containing exactly three leptons, possible ISR jets, and $\rm{E}_{\rm{T}}^{\rm{miss}}$.  The exclusion bound of $m_{\tilde{\chi}_2^0}$ was $640~\rm{GeV}$ for a massless $\tilde{\chi}_1^0$ in the wino scenario of the simplified model. It was weakened as the mass difference between $\tilde{\chi}_2^0$ and $\tilde{\chi}_1^0$ reduced. Specifically, $\tilde{\chi}_2^0$ should be heavier than $500~\rm{GeV}$ for $m_{\tilde{\chi}_1^0} = 300~{\rm GeV}$ (the on-shell W/Z case), $300~\rm{GeV}$ for a positive $m_{\tilde{\chi}_1^0}$ and $ 35~{\rm GeV} \lesssim m_{\tilde{\chi}_2^0} - m_{\tilde{\chi}_1^0} \lesssim 90~{\rm GeV}$ (the off-shell W/Z case), and $220~\rm{GeV}$ when $m_{\tilde{\chi}_1^0}$ is positive and $ m_{\tilde{\chi}_2^0} - m_{\tilde{\chi}_1^0} = 15~{\rm GeV}$ (the extreme off-shell W/Z case). By contrast, $\tilde{\chi}_2^0$ was excluded only up to a mass of $210~\rm{ GeV}$ in the off-shell W/Z case of the higgsino scenario, occurring when $ m_{\tilde{\chi}_2^0} - m_{\tilde{\chi}_1^0} = 10~{\rm GeV}$ or  $ m_{\tilde{\chi}_2^0} - m_{\tilde{\chi}_1^0} \gtrsim 35~{\rm GeV}$.

\item \texttt{ATLAS-1911-12606~\cite{ATLAS:2019lng}}: Search for the electroweakino pair productions and the slepton pair productions by two leptons and missing transverse momentum in the final state. This analysis concentrated on the scenario of compressed mass spectra and projected its results onto $\Delta m-\tilde{\chi}_2^0$ plane, where $\Delta m \equiv m_{\tilde{\chi}_2^0} -  m_{\tilde{\chi}_1^0}$ for the electroweakino production. It was found that the tightest bounds on the higgsino- and wino-dominated $\tilde{\chi}_2^0$ were $193~{\rm GeV}$ in mass for $\Delta m \simeq 9.3~{\rm GeV}$ and $240~{\rm GeV}$ in mass for $\Delta m \simeq 7~{\rm GeV}$, respectively. Similar lower mass limit on degenerate light-flavor sleptons was 250 GeV for $\Delta m_{\tilde{\ell}} \equiv m_{\tilde{\ell}} - m_{\tilde{\chi}_1^0} = 10~{\rm GeV}$.
\end{itemize}

\begin{table}[tbp]
\caption{The parameter space explored in this research, where $M_{\tilde{\ell}_L}$ and $M_{\tilde{\ell}_R}$ with $\ell = e, \mu$ denote the soft-breaking mass of the left- and right-handed slepton fields, respectively. The soft trilinear coefficients for the third-generation squarks, represented by $A_t$ and $A_b$, were assumed equal. The gluino mass was fixed at $M_3 =3~{\rm TeV}$. The other dimensional SUSY parameters were not crucial, and they took a shared value of $2~{\rm TeV}$ in this study, including $A_\lambda$ and the unmentioned soft-breaking parameters in the squark and slepton sectors. All the input parameters were defined at the renormalization scale $Q = 1~{\rm TeV}$.
\label{tab:3}}
\centering
\vspace{0.3cm}
\resizebox{0.7\textwidth}{!}{
\begin{tabular}{l|c|c||l|c|c}
\hline
Parameter & Prior & Range & Parameter & Prior & Range   \\
\hline
$\lambda$ & Flat & $ 0.01\sim0.7$ & $\kappa$ & Flat & $-0.7\sim0.7$ \\
$\tan \beta$ & Flat & $1.0 \sim60.0$ & $A_t/{\rm TeV}$ & Flat & $-5.0\sim5.0$ \\
$\mu/{\rm TeV}$ & Log & $~0.1\sim1.0$ &$A_\kappa/{\rm TeV}$ & Flat & $-1.0 \sim 1.0 $ \\
$M_1/{\rm TeV}$ & Flat & $-1.5\sim1.5$ & $M_2/{\rm TeV}$ & Log & $~0.1\sim1.5$ \\
$M_{\tilde{\mu}_L}/{\rm TeV}$ & Log & $0.1 \sim 1.0 $ &$M_{\tilde{\mu}_R}/{\rm TeV}$ & Log & $0.1 \sim 1.0$ \\
$M_{\tilde{e}_L}/{\rm TeV}$ & Log & $0.1 \sim 1.0$ &$M_{\tilde{e}_R}/{\rm TeV}$ & Log & $0.1 \sim 1.0$ \\
\hline
\end{tabular}}
\end{table}


\section{Numerical Results}\label{sec:na}

This research used the package \textsf{SARAH\,4.14.3}~\cite{Staub:2008uz, Staub:2012pb, Staub:2013tta, Staub:2015kfa} to build the model file of the $\mathbb{Z}_3$-NMSSM,
the codes \textsf{SPheno\,4.0.4}~\cite{Porod:2003um, Porod:2011nf} and \textsf{FlavorKit}~\cite{Porod:2014xia} to generate particle mass spectra and calculate low energy
observables such as $a_\ell^{\rm SUSY}$ and $B$-physics measurements,  and the package \textsf{MicrOMEGAs\,5.0.4}~\cite{Belanger:2001fz, Belanger:2005kh, Belanger:2006is, Belanger:2010pz, Belanger:2013oya, Barducci:2016pcb} to compute DM observables, assuming the lightest neutralino as the sole DM candidate in the universe. Bounds from the direct search for extra Higgs bosons at LEP, Tevatron, and LHC and the fit of Higgs property to pertinent experimental data were implemented by the codes~\textsf{HiggsBounds\,5.3.2}~\cite{HB2008jh,HB2011sb,HB2013wla,HB2020pkv} and \textsf{HiggsSignal\,2.2.3}~\cite{HS2013xfa,HSConstraining2013hwa,HS2014ewa,HS2020uwn}, respectively.

\subsection{Research strategy}

In our previous work~\cite{Cao:2022htd}, we performed two independent scans of the parameter space in the  $\mathbb{Z}_3$-NMSSM to reveal the salient features of $a_\mu^{\rm SUSY}$, using the \textsf{MultiNest} algorithm~\cite{Feroz:2008xx} with the setting $n_{\rm live}=8000$. In the first one, we fixed $A_\lambda = 2~{\rm TeV}$ and varied the parameters in the muon-type slepton, neutralino, and Higgs sectors. The second scan was similar to the first one, except we also varied $A_\lambda$. The results of these scans were identical in many aspects, indicating that the involved physics was insensitive to the parameter $A_\lambda$ or, more basically, the mass of the heavy doublet-dominated Higgs bosons, $m_A$, defined by $m_A^2 \equiv 2 \mu ( A_\lambda + 2 \kappa v_s)/\sin 2\beta$. In this research, we updated the first scan by varying the parameters in both the electron- and muon-type slepton sectors since we intended to explain the two leptonic anomalies simultaneously. We presented the details of the surveyed parameter space in Table~\ref{tab:3}, and correspondingly, we used the following likelihood function to guide the scan:
\begin{eqnarray}
\mathcal{L} &=& \mathcal{L}_{\rm other} \times \mathcal{L}_{a_{\mu}^{\rm SUSY}}\times \mathcal{L}_{a_e^{\rm SUSY}} \nonumber \\
&=& \mathcal{L}_{\rm other} \times \exp\left\{ -\frac{1}{2}\left [ \left( \frac{a_{\mu}^{\rm SUSY}- 2.51\times 10^{-9}}{5.9\times 10^{-10} }\right)^2 + \left( \frac{a_{e}^{\rm SUSY}- 4.8\times 10^{-13}}{3.0\times 10^{-13} }\right)^2\right] \right \}, \nonumber
\end{eqnarray}
where $\mathcal{L}_{\rm other}$ represented the impacts of pertinent experimental restrictions on the theory: $\mathcal{L}_{\rm other} = 1$ by our definition if the limitations are satisfied, and otherwise, $\mathcal{L}_{\rm other} = {\rm Exp[−100]}$. These limitations include:

\begin{itemize}
	\item DM relic density, $0.096 < \Omega h^2 < 0.144$. We took the central value of $\Omega {h^2}=0.120$ from the Planck-2018 data~\cite{Planck:2018vyg} and assumed a theoretical uncertainty of $20\%$ in the density calculation.
	\item DM direct detection bound from the XENON-1T experiments~\cite{Aprile:2018dbl,Aprile:2019dbj} on both the spin-independent (SI) DM-nucleon scattering cross-section, $\sigma_p^{\rm SI}$, and the spin-dependent (SD) neutron-nucleon cross-section, $\sigma_n^{\rm SD}$. The DM indirect detections from the observation of dwarf galaxies by the Fermi-LAT collaboration were not included since they had no restrictions on the theory when $|m_{\tilde{\chi}_1^0}| \gtrsim 100~{\rm GeV}$~\cite{Fermi-LAT:2015att}.
	\item Higgs data fit. Given that one of the CP-even Higgs bosons corresponded to the LHC-discovered Higgs boson, its properties should coincide with the Higgs measurements by ATLAS and CMS collaborations at the $95\%$ confidence level. A p-value larger than 0.05 is essential, which was tested by the code \textsf{HiggsSignal\, 2.2.3}~\cite{HS2013xfa,HSConstraining2013hwa,HS2014ewa,HS2020uwn}.
    \item Direct search for extra Higgs bosons at LEP, Tevatron and LHC. This requirement was examined by the code \textsf{HiggsBounds\, 5.3.2}~\cite{HB2008jh,HB2011sb,HB2013wla,HB2020pkv}.
	\item $B$-physics measurements. The branching ratios of $B_s \to \mu^+ \mu^-$ and $B \to X_s \gamma$ should be consistent with their experimental measurements at the $2\sigma$ level~\cite{PhysRevD.98.030001}.
	\item Vacuum stability. The vacuum state of the scalar potential comprising the Higgs fields and the first two-generation slepton fields should be either stable
or long-lived. This condition was tested by the code \textsf{Vevacious}~\cite{Camargo-Molina:2013qva}.
\end{itemize}

In this research, we were particularly interested in the samples that predicted $a_e^{\rm SUSY} >0$, explained the muon \texorpdfstring{$g-2$}{} anomaly at the $2 \sigma$ level, and meanwhile, coincided with all the restrictions.  We decided whether they pass the restrictions from the LHC search for SUSY in Tables~\ref{Table1} and~\ref{Table2} by simulating the following processes with the Monte Carlo (MC) method:
\begin{equation}
\begin{split}
pp \to \tilde{\chi}_i^0\tilde{\chi}_j^{\pm} &, \quad i = 2, 3, 4, 5; \quad j = 1, 2 \\
pp \to \tilde{\chi}_i^{\pm}\tilde{\chi}_j^{\mp} &, \quad i,j = 1, 2; \\
pp \to \tilde{\chi}_i^{0}\tilde{\chi}_j^{0} &, \quad i,j = 2, 3, 4, 5; \\
pp \to \tilde{\ell}_i^\ast \tilde{\ell}_j &,\quad i,j = L, R; \\
pp \to \tilde{\nu}_\ell^\ast \tilde{\nu}_\ell.
\end{split}
\end{equation}
Specifically, we calculated their cross-sections at $\sqrt{s}=13~{\rm TeV}$ by the package \textsf{Prospino2} \cite{Beenakker:1996ed} to the next-to
leading order. To save computation time, we initially used the program \textsf{SModelS\,2.1.1}, which encoded various event-selection efficiencies by the topologies of SUSY signals~\cite{Khosa:2020zar}, to exclude these samples. Given that this program's capability to implement the LHC restrictions was limited by its database and strict working prerequisites, we further surveyed the remaining samples by simulating the analyses in Tables~\ref{Table1} and~\ref{Table2}. We accomplished this task by following steps: we first generated 60000 and 40000 events for the electroweakino and slepton production processes, respectively, by the package \textsf{MadGraph\_aMC@NLO}~\cite{Alwall:2011uj, Conte:2012fm}, then finished the parton shower and hadronization by the program \textsf{PYTHIA8}~\cite{Sjostrand:2014zea}, and finally fed the resulting event files into the package \textsf{CheckMATE\,2.0.29}~\cite{Drees:2013wra,Dercks:2016npn, Kim:2015wza}, which incorporated the program \textsf{Delphes} for detector simulation~\cite{deFavereau:2013fsa}, to calculate the $R$-value defined by $R \equiv max\{S_i/S_{i,obs}^{95}\}$ ($S_i$ denotes the simulated event number of the $i$-th SR in the analyses of Tables~\ref{Table1} and~\ref{Table2}, and $S_{i,obs}^{95}$ represents its corresponding $95\%$ confidence level upper limit). Evidently, without considering the involved experimental and theoretical uncertainties, $R > 1$ implied that the studied parameter point was excluded due to its inconsistency with the LHC results~\cite{Domingo:2018ykx}. Otherwise, it was experimentally allowed.

\subsection{Key features of the results}

As we introduced before, this research strategy is the same as that of our previous work on the muon \texorpdfstring{$g-2$}{} anomaly in Ref.~\cite{Cao:2022htd}, except that we additionally included
the electron \texorpdfstring{$g-2$}{} anomaly in the likelihood function and correspondingly, we varied the parameters $M_{\tilde{e}_L}$ and $M_{\tilde{e}_R}$. Consequently, the two studies share many features in their results, which are summarized as follows:
\begin{itemize}
\item The DM candidate and the LHC-discovered Higgs boson are identified as the bino-dominated lightest neutralino and the lightest CP-even Higgs boson, respectively, in most cases. The Bayesian evidence of different scenarios testified to this conclusion~\cite{Cao:2022htd}.

\item The DM candidate $\tilde{\chi}_1^0$  achieves the measured relic density by co-annihilating with the wino-like electroweakinos or the sleptons.
Because the electroweakinos comprise two particles with approximately degenerate masses and both of them have stronger interactions with the SM particles than the sleptons, the former mechanism is easier to acquire the density~\cite{Cao:2022htd}. By contrast, the $Z$ and Higgs resonant annihilations are disfavored experimentally and theoretically in obtaining the density~\cite{He:2023lgi}. We will discuss this issue later.

\item Since a negative $\mu M_1$ can suppress the DM-nucleon SI scattering cross-section by cancelling different contributions, most samples yielded by the scan predict $M_1 < 0$~\cite{Cao:2022htd}. In addition, the XENON-1T experiments alone require $\mu \gtrsim 300~{\rm GeV}$, given the measured density~\cite{He:2023lgi}.

\item $\tilde{\chi}_1^0$ must be lighter than $620~{\rm GeV}$ to explain the $(g-2)_\mu$ anomaly at the $2\sigma$ level. With the increase of $|m_{\tilde{\chi}_1^0}|$, the maximum reach of $\mu$, $m_{\tilde{\mu}_L}$, and  $m_{\tilde{\mu}_R}$ decreases monotonously. This trend is more significant for $\mu$ and $m_{\tilde{\mu}_L}$ than for  $m_{\tilde{\mu}_R}$~\cite{Cao:2022htd}. The fundamental reason comes from the facts that the $\mathbb{Z}_3$-NMSSM is a decoupled theory in the heavy sparticle limit and $a_\mu^{\rm SUSY}$ is more sensitive to $\mu$ and  $m_{\tilde{\mu}_L}$ than to  $m_{\tilde{\mu}_R}$. $m_{\tilde{e}_L}$ and $m_{\tilde{e}_R}$ show similar behaviors when fixing $a_e^{\rm SUSY}$ at a positively sizable value.

\item Since explaining the muon \texorpdfstring{$g-2$}{} anomaly needs more than one sparticle to be moderately light, the restrictions from the LHC search for SUSY on the $\mathbb{Z}_3$-NMSSM are tight. As a result, the involved sparticles must be relatively heavy to coincide with the experimental results, e.g., $|m_{\tilde{\chi}_1^0}| \gtrsim 275~{\rm GeV}$, $m_{\tilde{\chi}_1^\pm} \gtrsim 300~{\rm GeV}$, and $m_{\tilde{\mu}_L} \gtrsim 310~{\rm GeV}$~\cite{Cao:2022htd}.
    The primary reasons are as follows: if $\tilde{\chi}_1^0$ is lighter, more missing momentum will be emitted in the sparticle production processes at the LHC, which can improve the sensitivities of the experimental analyses; if the sparticles other than $\tilde{\chi}_1^0$ are lighter, they will be copiously produced at the LHC to increase the events containing the multiple leptons.

    In addition, there are two cases that the LHC restrictions are extreme~\cite{Cao:2022htd,He:2023lgi}. One is characterized by $\tan \beta \lesssim 20$, where winos, higgsinos, and lefthanded dominant smuon are all lighter than 500 GeV to explain the muon \texorpdfstring{$g-2$}{} anomaly at the $2\sigma$ level. The other is characterized by predicting a $\tilde{\mu}_L$ lighter than winos and/or higgsinos, where the heavy electroweakinos may decay into $\tilde{\mu}_L$ first and thus enhance the leptonic signal of the electroweakino pair production processes (compared with the very massive $\tilde{\mu}_L$ case).

     We found that the signal regions of more than three leptons in CMS-SUS-16-039 and more than 200 GeV of $\rm{E}_{\rm{T}}^{\rm{miss}}$ in CMS-SUS-20-001 played a crucial role in excluding the parameter space~\cite{Cao:2022htd}. We presented more details of the LHC restrictions in Ref.~\cite{He:2023lgi}.

\item As required by the Higgs data fit, all the samples yielded by the scans predicted $\lambda \lesssim 0.3$. Consequently, the involved physics of the $\mathbb{Z}_3$-NMSSM is approximately the same as that of the MSSM. This conclusion was recently testified to by an elaborate study of the MSSM in Ref.~\cite{He:2023lgi}.

\item Since some of the involved sparticles can not be hefty, e.g., $\tilde{\chi}_1^0$ and $\tilde{\chi}_1^\pm$ should be lighter than $700~{\rm GeV}$, future colliders, such as the International Linear Collider with $\sqrt{s} = 1~{\rm TeV}$~\cite{ILC:2013jhg}, can explore different SUSY explanations of the anomalies~\cite{Chakraborti:2021squ}.
\end{itemize}

\begin{figure}[t]
	\centering
	\includegraphics[width=0.45\textwidth]{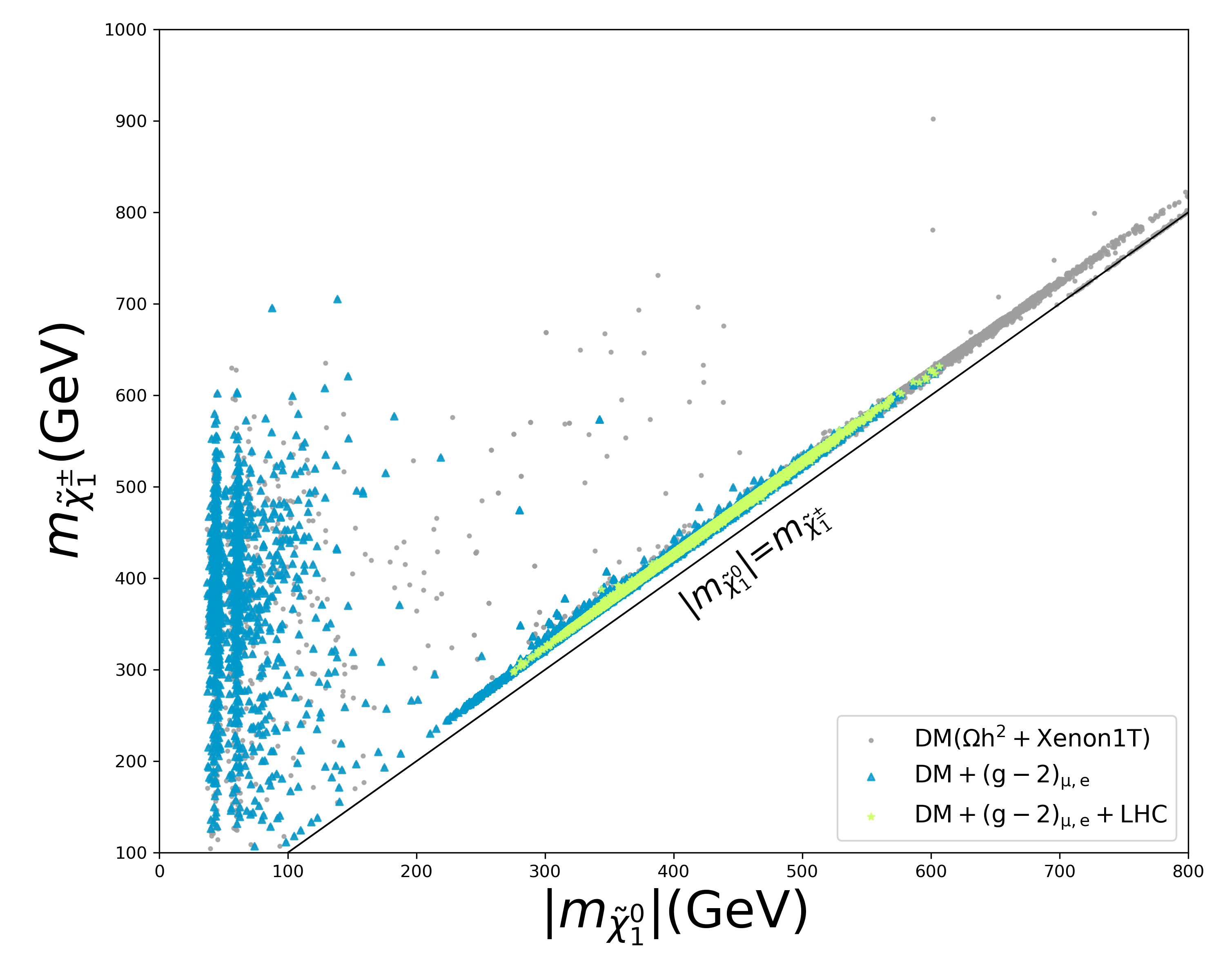}
	\includegraphics[width=0.45\textwidth]{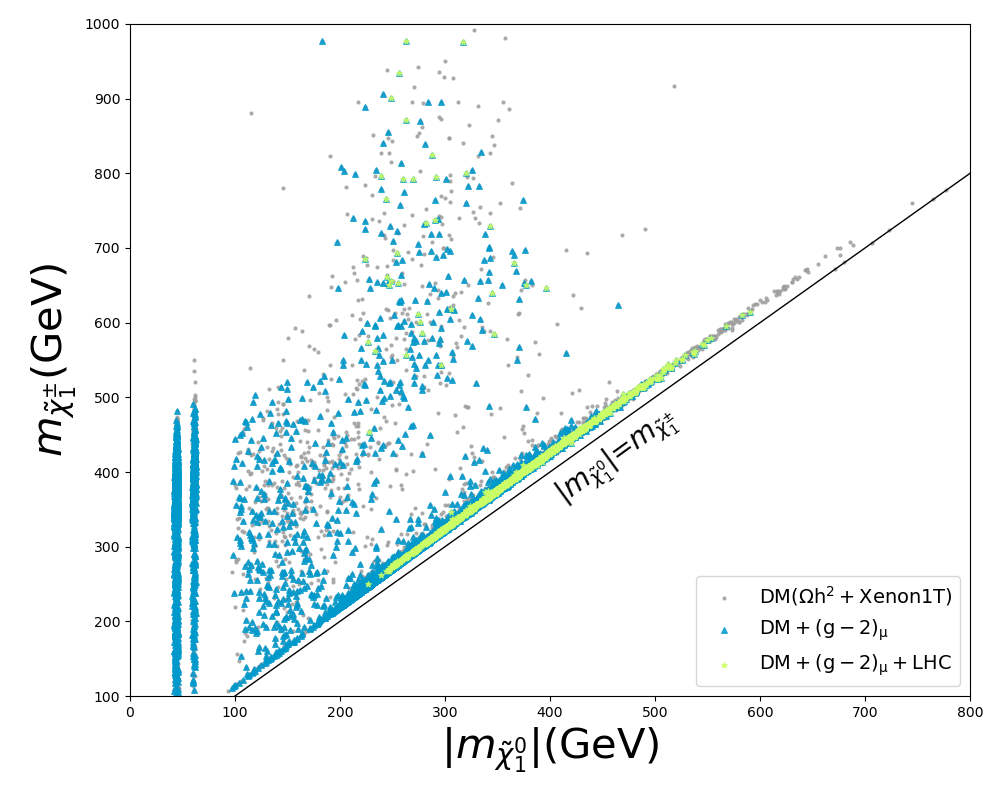}
	\caption{\label{fig1} Projection of the samples yielded by the first scan of Ref.~\cite{Cao:2022htd} and this research onto the $|m_{\tilde{\chi}_1^0}|-m_{\tilde{\chi}_1^\pm}$ planes, which are shown in the left and right panels, respectively. The grey dots denote the samples that survive all the restrictions listed in the text, in particular those from the DM experiments. The blue triangles represent the parameter points that predict $a_e^{\rm SUSY} > 0$ and further explain the muon \texorpdfstring{$g-2$}{} anomaly at the $2\sigma$ level. The green stars are part of the blue triangles which coincide with the results of the LHC search for SUSY.
}
\end{figure}

\begin{table}[]
\centering
	\caption{\label{Table4} The numbers of the blue and green samples in Fig.~\ref{fig1}, classified by the DM annihilation mechanisms. }	
	\vspace{0.2cm}

	\resizebox{0.95\textwidth}{!}{
\begin{tabular}{c|cc|cc}
\hline
\multicolumn{1}{l|}{Annihilation Mechanism}       & \multicolumn{2}{c|}{Numbers of the blue samples} & \multicolumn{2}{c}{Numbers of the green samples}   \\ \hline
                              & \multicolumn{1}{c|}{Left panel}  & \multicolumn{1}{c|}{Right panel}  & \multicolumn{1}{c|}{Left panel} & \multicolumn{1}{c}{Right panel} \\ \hline
\multicolumn{1}{l|}{All}                           & \multicolumn{1}{c|}{19178}         & \multicolumn{1}{c|}{19562}        & \multicolumn{1}{c|}{8736}        & \multicolumn{1}{c}{1398}        \\ \hline
\multicolumn{1}{l|}{Bino-Wino Co-annihilation}     & \multicolumn{1}{c|}{16415}         & \multicolumn{1}{c|}{15506}        & \multicolumn{1}{c|}{8627}        & \multicolumn{1}{c}{1327}        \\ \hline
\multicolumn{1}{l|}{Bino-Slepton Co-annihilation} & \multicolumn{1}{c|}{1514}         & \multicolumn{1}{c|}{2531}         & \multicolumn{1}{c|}{109}        & \multicolumn{1}{c}{71}          \\ \hline
\multicolumn{1}{l|}{$Z$-funnel}                    & \multicolumn{1}{c|}{691}         & \multicolumn{1}{c|}{993}          & \multicolumn{1}{c|}{0}        & \multicolumn{1}{c}{0          } \\ \hline
\multicolumn{1}{l|}{$h$-funnel}                    & \multicolumn{1}{c|}{558}         & \multicolumn{1}{c|}{532}          & \multicolumn{1}{c|}{0}        & \multicolumn{1}{c}{0}           \\ \hline
\end{tabular}}
\end{table}

Despite these similarities, there are two significant differences of the results, which include
\begin{itemize}
\item The posterior distributions of the samples yielded by the scans. To illustrate this point, we projected the SUSY parameter points acquired in the first scan of Ref.~\cite{Cao:2022htd} and those caught in this study onto the $|m_{\tilde{\chi}_1^0}|-m_{\tilde{\chi}_1^\pm}$ planes to obtain the left and right panels of Fig.~\ref{fig1}, respectively. The left panel indicates that nearly all the samples predicting $|m_{\tilde{\chi}_1^0}| \gtrsim 250~{\rm GeV}$ acquire the measured density by co-annihilating with the wino-like electroweakinos. By contrast, the right panel shows that a sizable portion of such points can achieve the measured density by co-annihilating with either the electron-type or muon-type sleptons. In addition, only sparse samples predict $ 100~{\rm GeV} \leq |m_{\tilde{\chi}_1^0}| \leq 250~{\rm GeV}$ in the left panel, while these samples are numerous in the right board. The primary reason for these distinctions is that the electron \texorpdfstring{$g-2$}{} anomaly prefers the involved particles, such as the wino-like and higgsino-like electroweakinos and $\tilde{e}_L$, to be lighter than the muon \texorpdfstring{$g-2$}{} anomaly.

  It is remarkable that the bound of $|m_{\tilde{\chi}_1^0}| \gtrsim 275~{\rm GeV}$, set by the LHC search for SUSY, in the left panel of Fig.~\ref{fig1} is significantly higher than that of $|m_{\tilde{\chi}_1^0}| \gtrsim 224~{\rm GeV}$ in the right board. This conclusion seems unreasonable since the LHC restrictions are tighter for the right panel (see the following discussions). The reason is that we studied the restrictions by simulating the signals of the samples acquired in the scan. So these lower bounds depend on the obtained parameter points or, more basically, their posterior distribution. Given the points are sparsely distributed in the low $|m_{\tilde{\chi}_1^0}|$ region of the left panel, the studied situations need to be more comprehensive to obtain an accurate bound. To testify to this speculation, we took the prior of $M_1$ in Table~\ref{Table1} to be log distributed and repeated the study of Ref.~\cite{Cao:2022htd}. We accumulated lots of samples in the low $|m_{\tilde{\chi}_1^0}|$ region and obtained a bound of $|m_{\tilde{\chi}_1^0}| \gtrsim 220~{\rm GeV}$ by concrete MC simulations. This bound came from the experimental analyses in Fig.~16 of Ref.~\cite{ATLAS:2021moa}, which concluded no LHC restrictions on winos in the $\tilde{B}-\tilde{W}$ co-annihilation case if $m_{\tilde{\chi}_1^0} \gtrsim 220~{\rm GeV}$. It is a physical bound, independent of the posterior distributions. We add that the lower bound of $|m_{\tilde{\chi}_1^0}|$ yielded in this study is very close to this experimental limit.

\item The LHC restrictions on the SUSY explanation. In Table~\ref{Table4}, we listed the numbers of the samples plotted in Fig.~\ref{fig1}. We classified them by the DM annihilation mechanisms and whether they passed the LHC restrictions. These numbers are approximately proportional to the posterior probabilities of the corresponding samples, and they reflect the LHC's capability to exclude the SUSY parameter points. They indicate that the impacts of the LHC restrictions were more important for this research than for that in Ref.~\cite{Cao:2022htd}. This conclusion arises from two physical reasons. One is that the electron \texorpdfstring{$g-2$}{} anomaly prefers lighter electroweakinos than the muon \texorpdfstring{$g-2$}{} anomaly. The other is due to the additional presence of light electron-type sleptons, which are copiously produced at the LHC and thus enhance the SUSY signals.

\end{itemize}

\subsection{Impacts of the LZ experiment }

\begin{figure}[t]
	\centering
	\includegraphics[width=0.45\textwidth]{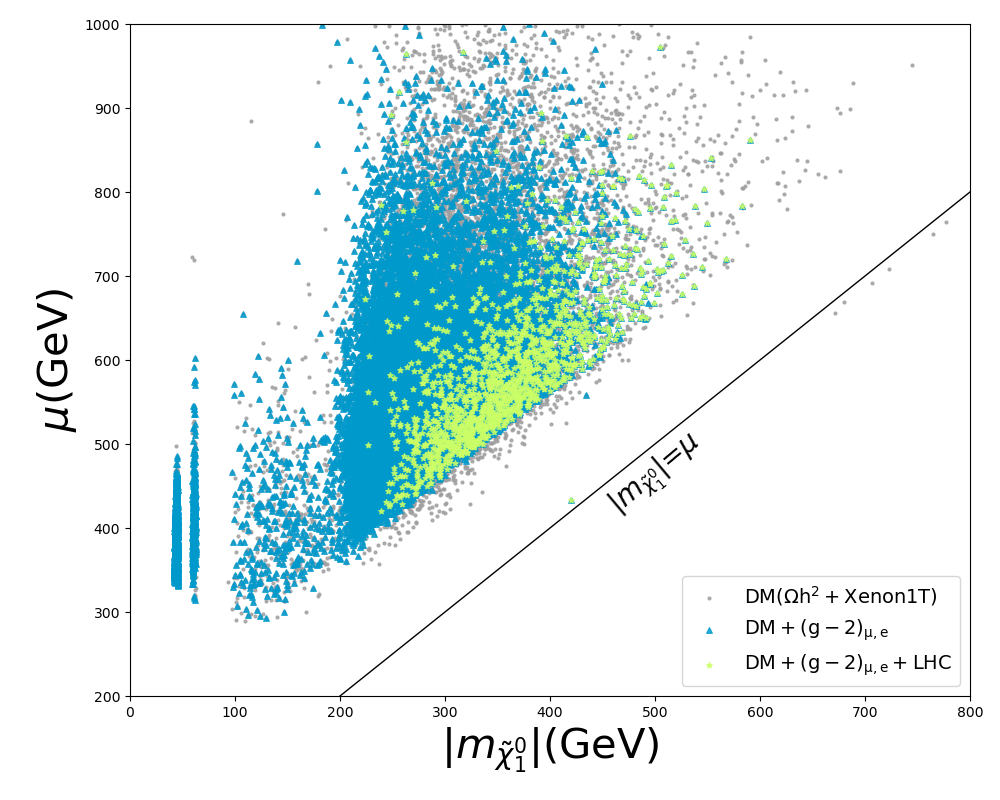}\hspace{-0.3cm}
	\includegraphics[width=0.45\textwidth]{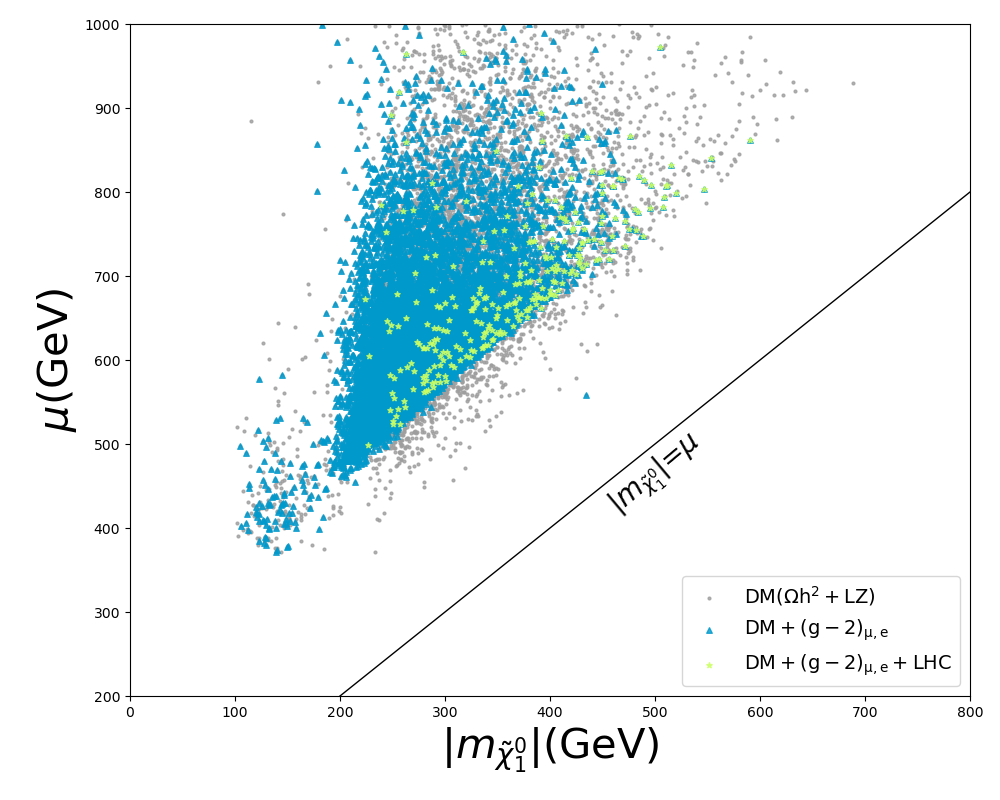}\hspace{-0.3cm}
	\caption{\label{fig2} Left panel: Same as the right panel of Fig.~\ref{fig1}, except that the samples are projected onto the $|m_{\tilde{\chi}^0_1}|-\mu$ plane. Right panel: Same as the left panel of this figure, except that all the parameter points further satisfy the LZ restrictions.  }
\end{figure}

\begin{table}[t]
	\centering
	\caption{\label{Table5} The numbers of the blue and green samples in the right panel of Fig.~\ref{fig2}, classified by the DM annihilation mechanisms. This table reflects the LHC's capability to exclude the SUSY parameter points that survive the LZ restrictions.}
	
	\vspace{0.2cm}
	\resizebox{0.95\textwidth}{!}{	
	\begin{tabular}{l|c|c}
		\hline
		Annihilation Mechanisms & Numbers of the blue samples &  Numbers of the green samples       \\ \hline
		\multicolumn{1}{l|}{All}                                                                          & 9180             & 277        \\
		\multicolumn{1}{l|}{Bino-Wino Co-annihilation}                                                                    & 8372              & 245        \\
		\multicolumn{1}{l|}{Bino-Sleptons Co-annihilation}                                                                                                   & 808                 &   32  \\
		\multicolumn{1}{l|}{$Z$-funnel}                                                                    & 0              &  0        \\
		\multicolumn{1}{l|}{$h$-funnel}                                                                    & 0              & 0         \\\hline
		
	\end{tabular}}
\end{table}

In the last section, we used the XENON-1T results to set upper limits on the SI and SD cross-sections of the DM-nucleon scattering~\cite{Aprile:2018dbl,Aprile:2019dbj}.
These restrictions, however, have been significantly improved by the recent LZ experiment~\cite{LZ:2022ufs}. Given this advancement, we refined the samples in the right panel of Fig.~\ref{fig1} with the LZ restrictions. After projecting the surviving parameter points onto various planes of SUSY parameters and comparing the resulting figures with their initial ones, we found the following three remarkable changes:
\begin{itemize}
\item Given the measured DM density, the LZ experiment alone required $\mu \gtrsim 380~{\rm GeV}$ for $|m_{\tilde{\chi}_1^0}| \simeq 100~{\rm GeV}$. This lower bound rose as $|m_{\tilde{\chi}_1^0}|$ increased, improving the corresponding XENON-1T restrictions by more than $50~{\rm GeV}$. This feature was shown in Fig.~\ref{fig2}.  Its primary reason was that  an enhanced $\mu$ could reduce the $\tilde{\chi}_1^0 \tilde{\chi}_1^0 Z$ and $\tilde{\chi}_1^0 \tilde{\chi}_1^0 h$ coupling strengthes and thus suppress the SI and SD scattering cross-sections~\cite{Cao:2019qng}.

\item As Fig.~\ref{fig2} of this research indicated, the LZ experiment made the $Z$-mediated resonant annihilation less preferred. The reason was that with the decrease of the $\tilde{\chi}_1^0 \tilde{\chi}_1^0 Z$ coupling strength, $2 |m_{\tilde{\chi}_1^0}|$ should be closer to $m_Z$ to obtain the measured density. This situation required the fine-tuning quantity defined in Eq. (19) of Ref.~\cite{Cao:2018rix} to be larger than 150. It usually led to the neglect of the resonant annihilation scenario in any less elaborated scan. This conclusion also applied to the $h$-mediated resonant annihilation. We add that this point was discussed by the Bayesian statistics in footnote 6 of Ref.~\cite{Cao:2022ovk}.

    One may rephrase this phenomenon as follows: since the relic density and the DM-nucleon scattering cross-sections are highly correlated in the $Z$- and $h$-funnel regions, great fine tunings of pertinent SUSY parameters are needed to acquire the measured density and satisfy the restrictions of the DM direct detection experiments simultaneously. This problem becomes more and more severe as the sensitivities of the direct detection experiments are improved.

\item Since the LZ and LHC restrictions were sensitive to different SUSY parameters, they complemented each other in probing the parameter space of the $\mathbb{Z}_3$-NMSSM. It was particularly so if one intended to explain the leptonic \texorpdfstring{$g-2$}{} anomalies at the $2 \sigma$ level, because $\mu$ was correlated with the other parameters by the anomalies, and any enhancement of $\mu$ in a massive higgsino scenario would make winos and $\tilde{\ell}_L$ lighter to keep $a_\ell^{\rm SUSY}$ unchanged. This situation usually tightens the LHC restrictions.

    In Table~\ref{Table5}, we listed the numbers of the parameter points in the right panel of Fig.~\ref{fig2}. Comparing them with the corresponding ones in the left panel of that figure, presented in Table~\ref{Table4}, one could readily conclude that, on the premise of explaining the leptonic anomalies at the $2\sigma$ level, the LZ experiment significantly promoted the LHC experiments to limit the parameter space of the $\mathbb{Z}_3$-NMSSM.

\end{itemize}

\begin{figure}[t]
	\centering
	\includegraphics[height=7cm,width=12cm]{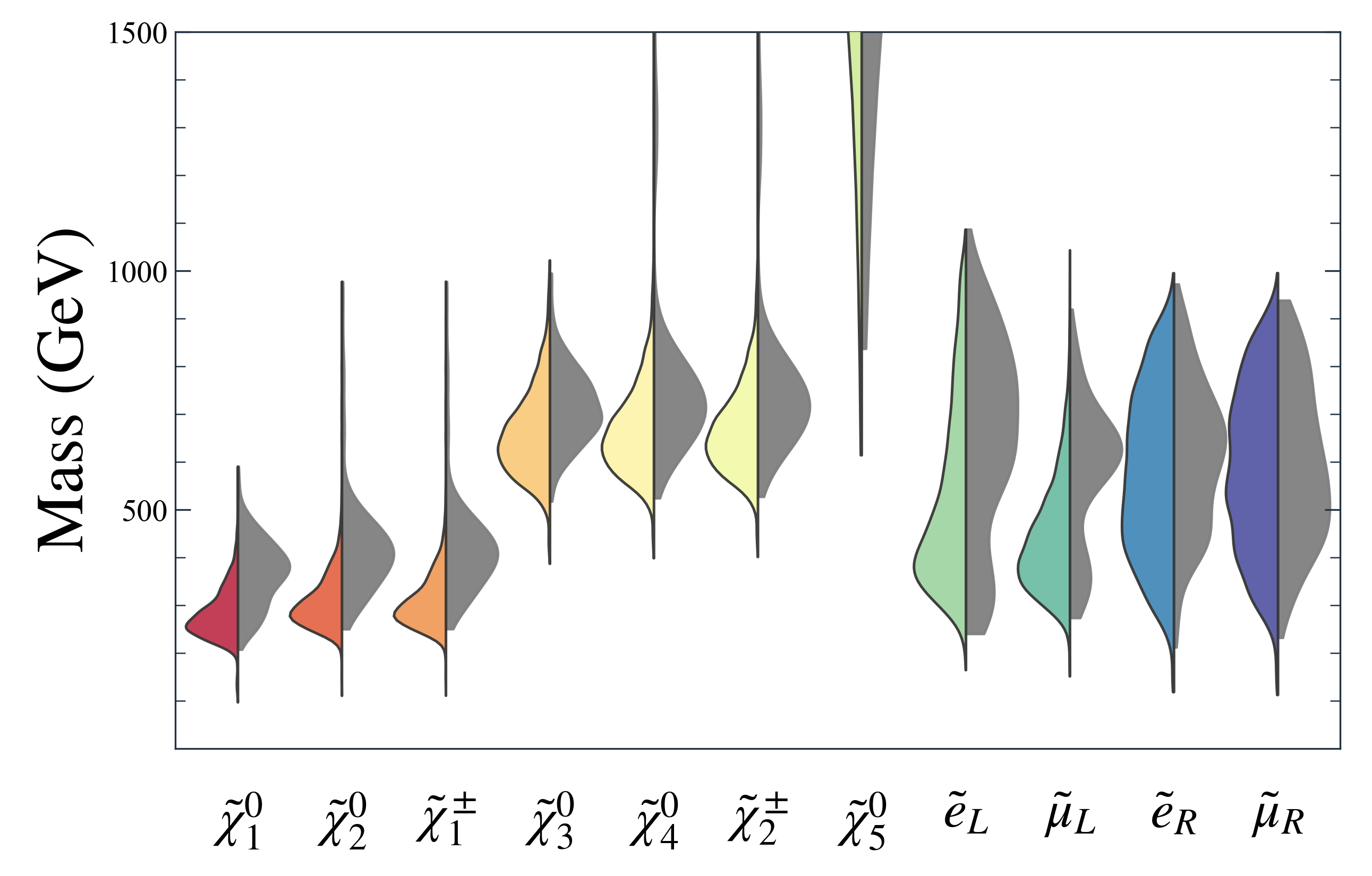} 
\vspace{-0.5cm}

	\caption{\label{fig3} Split violin plots with their shapes showing the mass distributions of the particles beyond the SM. This figure is plotted by counting the number of the samples yielded by the scan of this research with the LZ restrictions implemented. The left colorful side and the right gray side of each violin are based on 9188 and 277 samples, respectively. They are marked by the blue triangle and green star in the right panel of Fig.~\ref{fig2}. The widths of both sides are fixed so that their ratio does not represent the relative sample number. }
\end{figure}

Next, we focus on the sparticle mass spectra preferred by the leptonic anomalies, which were plotted in Fig.~\ref{fig3}. Through comparing the normalized mass distributions before and after including the LHC restrictions (corresponding to the left and right sides of the violin diagram for each particle in Fig.~\ref{fig3}), we conclude that the LHC restrictions affect little the shape of $m_{\tilde{\chi}_{5}^0}$ and $m_{\tilde{\mu}_R}$ distributions, but prefer more massive $\tilde{\chi}_{1}^0$, $\tilde{\chi}_{2}^0$, $\tilde{\chi}_{3}^0$, $\tilde{\chi}_1^\pm$, $\tilde{\chi}_2^\pm$, $\tilde{e}_L$, $\tilde{e}_R$, and $\tilde{\mu}_L$. The favored mass ranges for the latter set of particles are $ 600~{\rm GeV} \gtrsim m_{\tilde{\chi}_1^0} \gtrsim 224~{\rm GeV}$, $m_{\tilde{\chi}_2^0}, m_{\tilde{\chi}_1^\pm} \gtrsim 240~{\rm GeV}$, $m_{\tilde{\chi}_3^0} \gtrsim 510~{\rm GeV}$, $m_{\tilde{\chi}_4^0}, m_{\tilde{\chi}_2^\pm} \gtrsim 520~{\rm GeV}$, $m_{\tilde{\mu}_2} \gtrsim 250~{\rm GeV}$, $m_{\tilde{e}_L} \gtrsim 240~{\rm GeV}$, $m_{\tilde{e}_R} \gtrsim 230~{\rm GeV}$, $m_{\tilde{\mu}_L} \gtrsim 250~{\rm GeV}$, and $m_{\tilde{\mu}_R} \gtrsim 230~{\rm GeV}$, where the upper bound comes from the requirement to explain the leptonic \texorpdfstring{$g-2$}{} anomalies~\cite{Chakraborti:2020vjp}. We verified that $\tilde{\chi}_{2}^0$ and $\tilde{\chi}_1^\pm$ were wino-dominated when lighter than $500~{\rm GeV}$, and $\tilde{\chi}_5^0$ was always singlino-dominated. We emphasize that these ranges should be regarded as rough estimates rather than accurate bounds since the studied samples need more, given the broad parameter space of the $\mathbb{Z}_3$-NMSSM.

Finally, we have two comments on the acquired results:
\begin{itemize}
\item Throughout this research, we did not consider the theoretical uncertainties incurred by the simulations and the experimental (systematic and statistical) uncertainties. These effects could relax the LHC restrictions. However, given the advent of the high-luminosity LHC, much tighter restrictions on the $\mathbb{Z}_3$-NMSSM are expected to be available soon.

\item In some high-energy SUSY-breaking theories, $\tilde{\tau}$ may be the next-to-lightest supersymmetric particle. In this case, the signatures of the electroweakinos will significantly differ from those of this research. As a result, the LHC restrictions and, subsequently, the explanation of the anomalies may exhibit new features (see, e.g., the discussion in~\cite{Hagiwara:2017lse}). We will discuss such a possibility in the future.
\end{itemize}

\subsection{$a_e^{\rm SUSY}$ in the $\mathbb{Z}_3$-NMSSM}

\begin{figure}[t]
	\centering
	\includegraphics[width=0.45\textwidth]{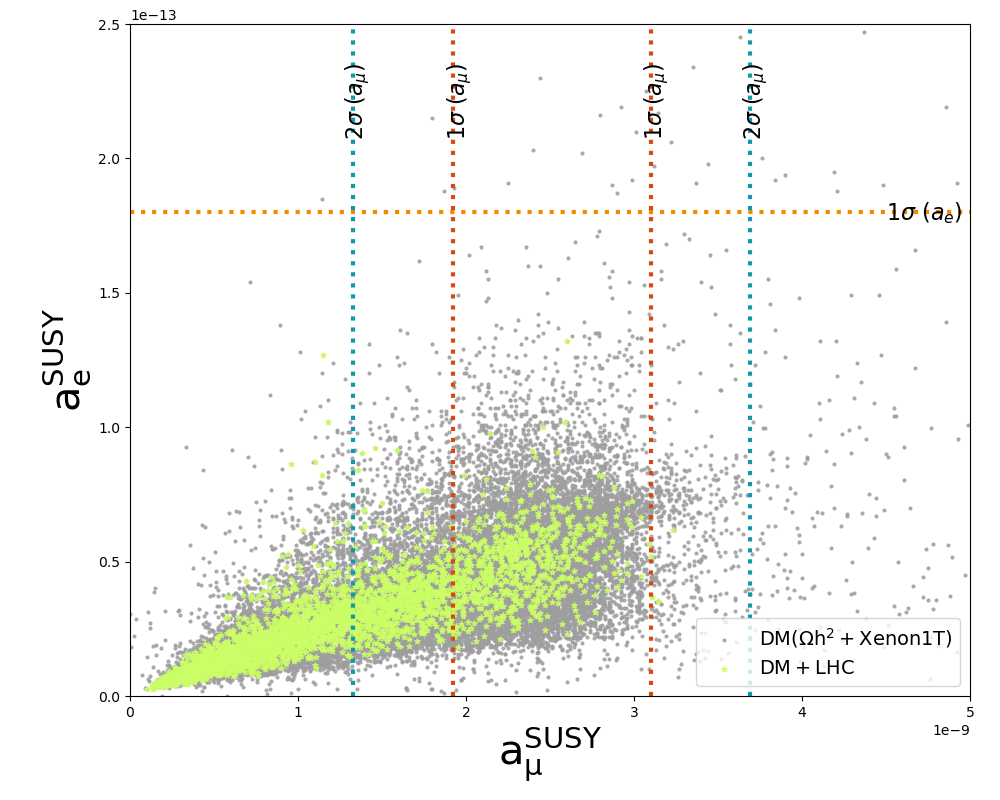}\hspace{-0.3cm}
	\includegraphics[width=0.45\textwidth]{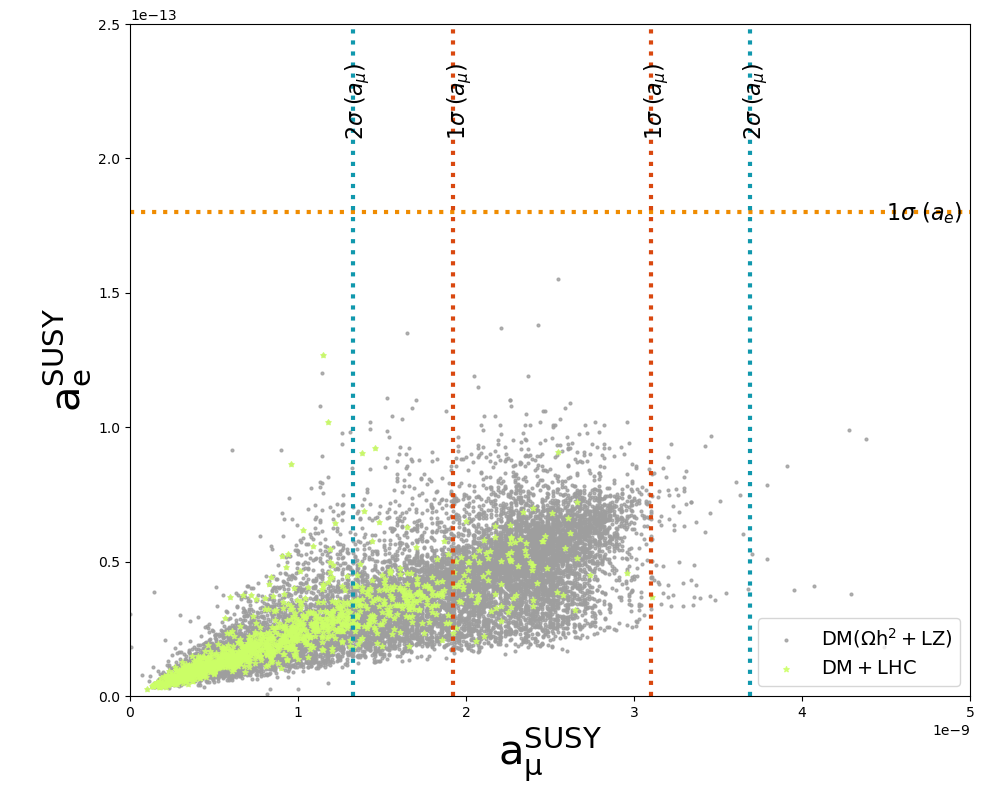}\hspace{-0.3cm}
	\caption{\label{fig4} Similar to Fig.~\ref{fig2}, except that the samples are projected onto the $a_{\mu}^{\rm SUSY}$ - $a_{ e}^{\rm SUSY}$ planes. The yellow dashed line corresponds to the $1 \sigma$ bound of $ a_{ e}^{\rm SUSY}$, and the red and blue dashed lines are the $1 \sigma$ and $2 \sigma$ bounds of $ a_{\mu}^{\rm SUSY}$, respectively. To acquire this figure, we specially included the samples predicting $a_e^{\rm SUSY} > 0 $ and $0 < a_\mu^{\rm SUSY} \leq 13.3 \times 10^{-10}$ and simulated their signals at the LHC. }
\end{figure}

\begin{table}[tpb]
	\caption{\label{tab:BP}Details of two benchmark points, P1 and P2. Both points predict $a_\mu^{\rm SUSY} \simeq 2.5 \times 10^{-9}$, but they correspond to significantly different $a_e^{\rm SUSY}$. }
	\vspace{0.3cm}
	\resizebox{1.0\textwidth}{!}{
		\begin{tabular}{llll|llll}
			\hline\hline
			\multicolumn{4}{l|}{\texttt{Benchmark Point P1}}& \multicolumn{4}{l}{\texttt{Benchmark Point P2}}                                                                        \\ \hline
			\multicolumn{1}{l}{$\lambda$}     & \multicolumn{1}{r}{0.151} & \multicolumn{1}{l}{$\text{m}_{\text{h}_s}$}    &\multicolumn{1}{r|}{961.7 GeV}     & \multicolumn{1}{l}{$\lambda$}     & \multicolumn{1}{r}{0.23} & \multicolumn{1}{l}{$\text{m}_{\text{h}_s}$}    &\multicolumn{1}{r}{2450.5 GeV}  \\
			\multicolumn{1}{l}{$\kappa$}  & \multicolumn{1}{r}{0.218} & \multicolumn{1}{l}{$\text{m}_{\text{A}_s}$}  &\multicolumn{1}{r|}{1166.4 GeV}     & \multicolumn{1}{l}{$\kappa$}     & \multicolumn{1}{r}{0.53} & \multicolumn{1}{l}{$\text{m}_{\text{A}_s}$}        &\multicolumn{1}{r}{914.3 GeV}     \\
			\multicolumn{1}{l}{$\text{tan}\beta$}      & \multicolumn{1}{r}{38.7} & \multicolumn{1}{l}{$\text{m}_\text{h}$}         &\multicolumn{1}{r|}{124.7 GeV}     & \multicolumn{1}{l}{$\text{tan}\beta$}    & \multicolumn{1}{r}{28.9} & \multicolumn{1}{l}{$\text{m}_\text{h}$}         &\multicolumn{1}{r}{124.7 GeV}   \\
			\multicolumn{1}{l}{$\mu$}     & \multicolumn{1}{r}{407.7 GeV} & \multicolumn{1}{l}{$\text{m}_{\text{A}_\text{H}}$}         &\multicolumn{1}{r|}{6258.3 GeV}     & \multicolumn{1}{l}{$\mu$}        & \multicolumn{1}{r}{541.4 GeV} & \multicolumn{1}{l}{$\text{m}_{\text{A}_\text{H}}$}         &\multicolumn{1}{r}{7047.0 GeV}     \\
			\multicolumn{1}{l}{$\text{A}_t$}        & \multicolumn{1}{r}{-3045.8 GeV} & \multicolumn{1}{l}{$\text{m}_{\tilde{\chi}_1^0}$}        &\multicolumn{1}{r|}{60.3 GeV}     & \multicolumn{1}{l}{$\text{A}_t$}        & \multicolumn{1}{r}{-2957.5 GeV} & \multicolumn{1}{l}{$\text{m}_{\tilde{\chi}_1^0}$}        &\multicolumn{1}{r}{-254.8 GeV}     \\
			\multicolumn{1}{l}{$\text{A}_{\kappa}$}        & \multicolumn{1}{r}{-776.7 GeV} & \multicolumn{1}{l}{$\text{m}_{\tilde{\chi}_2^0}$}        &\multicolumn{1}{r|}{175.5 GeV}     & \multicolumn{1}{l}{$\text{A}_{\kappa}$}        & \multicolumn{1}{r}{-231.0 GeV} & \multicolumn{1}{l}{$\text{m}_{\tilde{\chi}_2^0}$}        &\multicolumn{1}{r}{277.4 GeV}     \\
			\multicolumn{1}{l}{$\text{M}_1$}        & \multicolumn{1}{r}{61.3 GeV} & \multicolumn{1}{l}{$\text{m}_{\tilde{\chi}_3^0}$}        &\multicolumn{1}{r|}{-425.7 GeV}     & \multicolumn{1}{l}{$\text{M}_1$}        & \multicolumn{1}{r}{-256.3 GeV} & \multicolumn{1}{l}{$\text{m}_{\tilde{\chi}_3^0}$}        &\multicolumn{1}{r}{-561.1 GeV}     \\
			\multicolumn{1}{l}{$\text{M}_2$}        & \multicolumn{1}{r}{171.9 GeV} & \multicolumn{1}{l}{$\text{m}_{\tilde{\chi}_4^0}$}        &\multicolumn{1}{r|}{434.0 GeV}     & \multicolumn{1}{l}{$\text{M}_2$}        & \multicolumn{1}{r}{270.0 GeV} & \multicolumn{1}{l}{$\text{m}_{\tilde{\chi}_4^0}$}        &\multicolumn{1}{r}{566.9 GeV}     \\
			\multicolumn{1}{l}{$\text{M}_{\tilde{e}_{\text{L}}}$}        & \multicolumn{1}{r}{127.1 GeV} & \multicolumn{1}{l}{$\text{m}_{\tilde{\chi}_5^0}$}        &\multicolumn{1}{r|}{1176.5 GeV}     & \multicolumn{1}{l}{$\text{M}_{\tilde{e}_{\text{L}}}$}        & \multicolumn{1}{r}{124.8 GeV} & \multicolumn{1}{l}{$\text{m}_{\tilde{\chi}_5^0}$}        &\multicolumn{1}{r}{2521.3 GeV}     \\
			\multicolumn{1}{l}{$\text{M}_{\tilde{e}_{\text{R}}}$}        & \multicolumn{1}{r}{573.0 GeV} & \multicolumn{1}{l}{$\text{m}_{\tilde{\chi}_1^{\pm}}$}        &\multicolumn{1}{r|}{175.7 GeV}     & \multicolumn{1}{l}{$\text{M}_{\tilde{e}_{\text{R}}}$}        & \multicolumn{1}{r}{805.4 GeV} & \multicolumn{1}{l}{$\text{m}_{\tilde{\chi}_1^{\pm}}$}      &\multicolumn{1}{r}{277.6 GeV}     \\
			\multicolumn{1}{l}{$\text{M}_{\tilde{\mu}_{\text{L}}}$}        & \multicolumn{1}{r}{663.2 GeV} & \multicolumn{1}{l}{$\text{m}_{\tilde{\chi}_2^{\pm}}$}        &\multicolumn{1}{r|}{437.2 GeV}     & \multicolumn{1}{l}{$\text{M}_{\tilde{\mu}_{\text{L}}}$}        & \multicolumn{1}{r}{278.5 GeV} & \multicolumn{1}{l}{$\text{m}_{\tilde{\chi}_2^{\pm}}$}        &\multicolumn{1}{r}{570.0 GeV}     \\
			\multicolumn{1}{l}{$\text{M}_{\tilde{\mu}_{\text{R}}}$}        & \multicolumn{1}{r}{707.6 GeV} & \multicolumn{1}{l}{$\text{m}_{\tilde{e}_{\text{L}}}$}        &\multicolumn{1}{r|}{256.0 GeV}     & \multicolumn{1}{l}{$\text{M}_{\tilde{\mu}_{\text{R}}}$}        & \multicolumn{1}{r}{767.0 GeV} & \multicolumn{1}{l}{$\text{m}_{\tilde{e}_{\text{L}}}$}       &\multicolumn{1}{r}{287.3 GeV}     \\
			\multicolumn{1}{l}{$a_{e}^{\text{SUSY}}$}       & \multicolumn{1}{r}{$2.66 \times 10^{-13}$} & \multicolumn{1}{l}{$\text{m}_{\tilde{e}_{\text{R}}}$}        &\multicolumn{1}{r|}{487.7 GeV}     & \multicolumn{1}{l}{$a_{e}^{\text{SUSY}}$}       & \multicolumn{1}{r}{$6.8 \times 10^{-14}$} & \multicolumn{1}{l}{$\text{m}_{\tilde{e}_{\text{R}}}$}      &\multicolumn{1}{r}{725.4 GeV}     \\
			\multicolumn{1}{l}{$a_{\mu}^{\text{SUSY}}$}       & \multicolumn{1}{r}{$2.44 \times 10^{-9}$} & \multicolumn{1}{l}{$\text{m}_{\tilde{\mu}_{\text{L}}}$}        &\multicolumn{1}{r|}{700.6 GeV}     & \multicolumn{1}{l}{$a_{\mu}^{\text{SUSY}}$}       & \multicolumn{1}{r}{$2.51 \times 10^{-9}$} & \multicolumn{1}{l}{$\text{m}_{\tilde{\mu}_{\text{L}}}$}      &\multicolumn{1}{r}{381.2 GeV}     \\
			\multicolumn{1}{l}{${\Omega h}^2$}     & \multicolumn{1}{r}{0.14} & \multicolumn{1}{l}{$\text{m}_{\tilde{\mu}_{\text{R}}}$}      &\multicolumn{1}{r|}{641.1 GeV}     & \multicolumn{1}{l}{${\Omega h}^2$}     & \multicolumn{1}{r}{0.10} & \multicolumn{1}{l}{$\text{m}_{\tilde{\mu}_{\text{R}}}$}    &\multicolumn{1}{r}{683.0 GeV}     \\
			\multicolumn{1}{l}{$\sigma_{p}^{\text{SI}}$}   & \multicolumn{1}{r}{$3.40 \times 10^{-47}\text{cm}^{2}$} & \multicolumn{1}{l}{$\text{m}_{\tilde{\nu}_{e}}$}      &\multicolumn{1}{r|}{243.9 GeV}     & \multicolumn{1}{l}{$\sigma_{p}^{\text{SI}}$}   &\multicolumn{1}{r}{$5.98 \times 10^{-47}\text{cm}^{2}$}& \multicolumn{1}{l}{$\text{m}_{\tilde{\nu}_{e}}$}    &\multicolumn{1}{r}{276.7 GeV}     \\
			\multicolumn{1}{l}{$\sigma_{n}^{\text{SD}}$}   & \multicolumn{1}{r}{$3.95 \times 10^{-42}\text{cm}^2$} & \multicolumn{1}{l}{$\text{m}_{\tilde{\nu}_{\mu}}$}    &\multicolumn{1}{r|}{695.8 GeV}     & \multicolumn{1}{l}{$\sigma_{n}^{\text{SD}}$}   & \multicolumn{1}{r}{$1.74 \times 10^{-42}\text{cm}^2$} & \multicolumn{1}{l}{$\text{m}_{\tilde{\nu}_{\mu}}$}    &\multicolumn{1}{r}{373.3 GeV}     \\ \hline
			\multicolumn{2}{l}{$\text{N}_{11},\text{N}_{12},\text{N}_{13},\text{N}_{14},\text{N}_{15}$}           & \multicolumn{2}{l|}{-0.994,~~0.014,~-0.109,~~0.019,~-0.003}                 & \multicolumn{2}{l}{$\text{N}_{11},\text{N}_{12},\text{N}_{13},\text{N}_{14},\text{N}_{15}$}           & \multicolumn{2}{l}{-0.994,~-0.006,~-0.097,~-0.042,~-0.002}          \\
			\multicolumn{2}{l}{$\text{N}_{21},\text{N}_{22},\text{N}_{23},\text{N}_{24},\text{N}_{25}$}           & \multicolumn{2}{l|}{-0.039,~-0.970,~~0.219,~-0.096,~~0.006}                 & \multicolumn{2}{l}{$\text{N}_{21},\text{N}_{22},\text{N}_{23},\text{N}_{24},\text{N}_{25}$}           & \multicolumn{2}{l}{~0.009,~~0.978,~-0.187,~~0.098,~-0.003}                 \\
			\multicolumn{2}{l}{$\text{N}_{31},\text{N}_{32},\text{N}_{33},\text{N}_{34},\text{N}_{35}$}           & \multicolumn{2}{l|}{-0.062,~~0.090,~~0.697,~~0.709,~~0.011}                 & \multicolumn{2}{l}{$\text{N}_{31},\text{N}_{32},\text{N}_{33},\text{N}_{34},\text{N}_{35}$}           & \multicolumn{2}{l}{~0.098,~-0.064,~-0.698,~-0.706,~-0.009}                 \\
			\multicolumn{2}{l}{$\text{N}_{41},\text{N}_{42},\text{N}_{43},\text{N}_{44},\text{N}_{45}$}           & \multicolumn{2}{l|}{~0.084,~-0.225,~-0.674,~~0.699,~-0.023}                 & \multicolumn{2}{l}{$\text{N}_{41},\text{N}_{42},\text{N}_{43},\text{N}_{44},\text{N}_{45}$}           & \multicolumn{2}{l}{~0.038,~-0.201,~-0.684,~~0.700,~-0.013}       \\
			\multicolumn{2}{l}{$\text{N}_{51},\text{N}_{52},\text{N}_{53},\text{N}_{54},\text{N}_{55}$}           & \multicolumn{2}{l|}{-0.001,~-0.001,~-0.025,~~0.008,~~0.999}                 & \multicolumn{2}{l}{$\text{N}_{51},\text{N}_{52},\text{N}_{53},\text{N}_{54},\text{N}_{55}$}           & \multicolumn{2}{l}{-0.001,~-0.000,~-0.016,~~0.003,~~1.000}      \\ \hline
			\multicolumn{2}{l}{Annihilations}                     & \multicolumn{2}{l|}{Fractions[\%]}        & \multicolumn{2}{l}{Annihilations}                     & \multicolumn{2}{l}{Fractions[\%]}        \\
			\multicolumn{2}{l}{$\tilde{\chi}_1^0\tilde{\chi}_1^0 \to f\bar{f}/g g$}                  & \multicolumn{2}{l|}{82.8/17.2}                 & \multicolumn{2}{l}{$\tilde{B}-\tilde{W}$ Co-annihilation}            & \multicolumn{2}{l}{100}                 \\ \hline
			\multicolumn{2}{l}{Decays}                            & \multicolumn{2}{l|}{Branching ratios[\%]} & \multicolumn{2}{l}{Decays}                            & \multicolumn{2}{l}{Branching ratios[\%]} \\
			\multicolumn{2}{l}{$\tilde{\chi}_2^0 \to \tilde{\chi}_1^0Z$}& \multicolumn{2}{l|}{98.1}  &\multicolumn{2}{l}{$\tilde{\chi}_2^0 \to \tilde{\nu}_{e}\nu_{e}$}& \multicolumn{2}{l}{99.5}                 \\
			\multicolumn{2}{l}{$\tilde{\chi}_3^0 \to \tilde{\chi}_1^{\pm}W^{\mp}/\tilde{\chi}_2^0 Z/\tilde{\chi}_1^0 Z/\tilde{\chi}_1^0 h/\tilde{\chi}_2^0 h$} & \multicolumn{2}{l|}{62.1/23.8/7.4/3.3/2.9}                 & \multicolumn{2}{l}{$\tilde{\chi}_3^0 \to \tilde{\chi}_1^{\pm}W^{\mp}/\tilde{\chi}_2^0 Z/\tilde{\chi}_1^0h/\tilde{\chi}_2^0h/\tilde{\chi}_1^0Z$}                           & \multicolumn{2}{l}{61.6/26.4/7.4/2.2/1.4}                 \\
			\multicolumn{2}{l}{$\tilde{\chi}_4^0 \to \tilde{\chi}_1^{\pm}W^{\mp}/\tilde{\chi}_2^0 h/\tilde{\chi}_1^0 h/\tilde{\chi}_2^0 Z/\tilde{\chi}_1^0 Z/\tilde{\nu}_{e}\nu_{e}/\tilde{e}^{\pm}_{\text{L}}e^{\mp}$}& \multicolumn{2}{l|}{62.6/19.9/6.2/4.4/3.6/2.4/1.0}                 & \multicolumn{2}{l}{$\tilde{\chi}_4^0 \to \tilde{\chi}_1^{\pm}W^{\mp}/\tilde{\chi}_2^0 h/\tilde{\chi}_1^0 Z/\tilde{\chi}_2^0 Z/\tilde{\nu}_{e}\nu_{e}/\tilde{e}^{\pm}_{\text{L}} e^{\mp}/\tilde{\nu}_{\mu}\nu_{\mu}/\tilde{\chi}_1^0 h$}     & \multicolumn{2}{l}{60.6/23.3/7.1/3.0/1.9/1.2/1.1/1.1}                 \\
			\multicolumn{2}{l}{$\tilde{\chi}_5^0 \to \tilde{\chi}_2^{\pm}W^{\mp}/\tilde{\chi}_4^0 h/\tilde{\chi}_3^0 Z/\tilde{\chi}_3^0 h/\tilde{\chi}_1^{\pm}W^{\mp}//\tilde{\chi}_4^0 Z/\tilde{\chi}_2 h/\tilde{\chi}_2 Z$} & \multicolumn{2}{l|}{44.3/19.2/18.8/4.8/4.6/4.2/1.3/1.1}                 & \multicolumn{2}{l}{$\tilde{\chi}_5^0 \to \tilde{\chi}_2^{\pm}W^{\mp}/\tilde{\chi}_4^0 h/\tilde{\chi}_3^0 Z/\tilde{\chi}_3^0 h/\tilde{\chi}_4^0 Z/\tilde{\chi}_1^{\pm}W^{\mp}/\tilde{\chi}_2^0 h$}& \multicolumn{2}{l}{44.7/16.1/15.9/7.8/7.0/3.3/1.0}         \\	
			\multicolumn{2}{l}{$\tilde{\chi}_1^{\pm} \to \tilde{\chi}_1^0W^{\pm}$}   & \multicolumn{2}{l|}{99.4}                 & \multicolumn{2}{l}{$\tilde{\chi}_1^{\pm} \to \tilde{\nu}_{e} e^{\mp}$}                           & \multicolumn{2}{l}{99.8}                 \\
			\multicolumn{2}{l}{$\tilde{\chi}_2^{\pm} \to \tilde{\chi}_2^0 W^{\pm}/\tilde{\chi}_1^{\pm} Z/\tilde{\chi}_1^{\pm} h/\tilde{\chi}_1^0W^{\pm}/\tilde{e}_{\text{L}} \nu_{e}$}        & \multicolumn{2}{l|}{32.7/29.7/24.1/9.9/3.0}& \multicolumn{2}{l}{$\tilde{\chi}_2^{\pm} \to \tilde{\chi}_2^0W^{\pm}/\tilde{\chi}_1^{\pm}Z/\tilde{\chi}_1^{\pm} h/\tilde{\chi}_1^0W^{\pm}/\tilde{e}_{\text{L}}^{\pm} \nu_{e}/\tilde{\mu}_{\text{L}}^{\pm} \nu_{\mu}$}  & \multicolumn{2}{l}{30.6/29.2/25.6/9.3/2.6/1.5}                 \\
			\multicolumn{2}{l}{$\tilde{e}_{\text{L}}^{\pm} \to \tilde{\chi}_1^{\pm}\nu_{e}/\tilde{\chi}_2^0 e^{\pm}/\tilde{\chi}_1^0 e^{\pm}$} & \multicolumn{2}{l|}{48.2/25.4/26.4}& \multicolumn{2}{l}{$\tilde{e}_{\text{L}}^{\pm} \to \tilde{\chi}_1^0 e^{\pm}/\tilde{\chi}_1^{\pm}\nu_{e}/\tilde{\chi}_2^0 e^{\pm}$}  & \multicolumn{2}{l}{53.7/29.9/16.4}\\
			\multicolumn{2}{l}{$\tilde{e}_{\text{R}}^{\pm} \to \tilde{\chi}_1^0 e^{\pm}$}  & \multicolumn{2}{l|}{99.8} & \multicolumn{2}{l}{$\tilde{e}_{\text{R}}^{\pm} \to \tilde{\chi}_1^0 e^{\pm}$} &\multicolumn{2}{l}{99.7}\\
			\multicolumn{2}{l}{$\tilde{\mu}_{\text{L}}^{\pm} \to \tilde{\chi}_1^{\pm}\nu_{\mu}/\tilde{\chi}_2^0\mu^{\pm}/\tilde{\chi}_1^{0}\mu^{\pm}/\tilde{\chi}_2^{\pm}\nu_{\mu}$} & \multicolumn{2}{l|}{56.0/30.5/10.4/2.5}   & \multicolumn{2}{l}{$\tilde{\mu}_{\text{L}}^{\pm} \to \tilde{\chi}_1^{\pm}\nu_{\mu}/\tilde{\chi}_2^0\mu^{\pm}/\tilde{\chi}_1^0\mu^{\pm}$}  & \multicolumn{2}{l}{56.8/29.6/13.6} \\
			\multicolumn{2}{l}{$\tilde{\mu}_{\text{R}}^{\pm} \to \tilde{\chi}_1^0\mu^{\pm}$} & \multicolumn{2}{l|}{99.2}                 & \multicolumn{2}{l}{$\tilde{\mu}_{\text{R}}^{\pm} \to \tilde{\chi}_1^0\mu^{\pm}$} &\multicolumn{2}{l}{99.6}\\	
			\multicolumn{2}{l}{$\tilde{\nu}_{e} \to \tilde{\chi}_1^{\pm} e^{\mp}/\tilde{\chi}_1^{0}\nu_{e}/\tilde{\chi}_2^{0}\nu_{e}$}     & \multicolumn{2}{l|}{47.6/30.5/21.9}                 & \multicolumn{2}{l}{$\tilde{\nu}_{e} \to \tilde{\chi}_1^{\pm} e^{\mp}$}                             & \multicolumn{2}{l}{100}                 \\
			\multicolumn{2}{l}{$\tilde{\nu}_{\mu} \to \tilde{\chi}_1^{\pm}\mu^{\mp}/\tilde{\chi}_2^{0}\nu_{\mu}/\tilde{\chi}_1^{0}\nu_{\mu}/\tilde{\chi}_4^{0}\nu_{\mu}$}     	& \multicolumn{2}{l|}{59.7/27.3/11.3/1.0}                 & \multicolumn{2}{l}{$\tilde{\nu}_{\mu} \to \tilde{\chi}_1^{\pm}\mu^{\mp}/\tilde{\chi}_2^0\nu_{\mu}/\tilde{\chi}_1^{0}\nu_{\mu}$}                             & \multicolumn{2}{l}{58.5/28.3/13.2}   \\ \hline
			\multicolumn{2}{l}{$R$ value}   & \multicolumn{2}{l|}{4.83, ATLAS-SUSY-2018-06}                 & \multicolumn{2}{l}{$R$ value}                           & \multicolumn{2}{l}{0.75, CMS-SUS-20-001}                 \\ \hline\hline
	\end{tabular}}
\end{table}

To learn the typical size of $a_e^{\rm SUSY}$ in the $\mathbb{Z}_3$-NMSSM, we projected the parameter points yielded by the scan of this research onto the $a_{\mu}^{\rm SUSY}$ - $a_{ e}^{\rm SUSY}$ plane. The results were shown in Fig.~\ref{fig4}, where the samples in the left panel coincided with the XENON-1T results, while those in the right board further satisfied the LZ restrictions. This figure revealed the following features:
\begin{itemize}
\item The formula $a_e^{\rm SUSY}/m_e^2 \simeq a_\mu^{\rm SUSY}/m_\mu^2$ correlated $a_e^{\rm SUSY}$ and $a_\mu^{\rm SUSY}$ in most cases. The primary reason was that the two quantities depended similarly on the parameters $M_1$, $M_2$, and $\mu$, and meanwhile, the electron-type and muon-type sleptons were comparable in mass. Consequently, $a_e^{\rm SUSY}$ was typically around $5 \times 10^{-14}$ when $a_\mu^{\rm SUSY} \simeq 2.5 \times 10^{-9}$. In addition, we found the violations of this correlation in rare cases such as $m_{\tilde{\mu}_L} \gg M_2, \mu, m_{\tilde{e}_L}$, where $a_e^{\rm SUSY}/m_e^2$ could be much larger than $a_\mu^{\rm SUSY}/m_\mu^2$.

\item The DM direct detection experiments and the LHC experiments were tight in restricting the case of $a_e^{\rm SUSY} \gtrsim 1.0 \times 10^{-13}$ or $a_\mu^{\rm SUSY} \gtrsim 3.0 \times 10^{-9}$, where some sparticles involved in the leptonic anomalies should be relatively light, as revealed by the discussions of this research. This fact implied that $a_e^{\rm SUSY}$ and  $a_\mu^{\rm SUSY}$ could not be huge. Specifically, $a_e^{\rm SUSY}$ and $a_\mu^{\rm SUSY}$ might reach $3 \times 10^{-13}$ and $5 \times 10^{-9}$, respectively, in the optimum cases if one used the XENON-1T experiment to limit the SUSY parameter space. The predictions, however, reduced to $1.5 \times 10^{-13}$ and $4 \times 10^{-9}$ correspondingly after implementing the LZ restrictions and $1.0 \times 10^{-13}$ and $3 \times 10^{-9}$ when further considering the LHC restrictions.

    In Table~\ref{tab:BP}, we provided the details of two benchmark points, P1 and P2, both predicting $a_\mu^{\rm SUSY} \simeq 2.5 \times 10^{-9}$. Point P1 corresponded to a relatively large $a_e^{\rm SUSY}$, i.e., $a_e^{\rm SUSY} \simeq 2.7 \times 10^{-13}$. It satisfied the XENON-1T restrictions but was excluded by the LZ and LHC experiments. By contrast, $a_e^{\rm SUSY}$ of point P2 was moderate, namely $a_e^{\rm SUSY} = 6.8 \times 10^{-14}$, and this point coincided with all the experimental results.
\end{itemize}

\section{Summary}\label{sec:sum}

In recent years we have witnessed the rapid progress of particle physics experiments. In particular, the discrepancy between $a_\mu^{\rm Exp}$ and $a_\mu^{\rm SM}$ was corroborated by the E989 experiment at FNAL, and the significant departures of $a_e^{\rm Exp}$ from $a_e^{\rm SM}$ were also observed in the measurements of the fine structure constant at LKB and LBNL. These anomalies hinted at the existence of new physics, and SUSY as the most compelling new physics candidate has attracted a lot of attention.

So far, most of the research on the leptonic anomalies concentrated on the MSSM, which, however, suffered increasing fine-tuning problems as the smooth ongoing of the LHC experiments.
This situation motivated us to explore this subject in another economic and self-contained low-energy realization of SUSY, namely the $\mathbb{Z}_3$-NMSSM. Specifically, we investigated how large $a_e^{\rm SUSY}$ could reach in the $\mathbb{Z}_3$-NMSSM after considering various experimental restrictions, given that this quantity was scarcely studied in SUSY. For this purpose,  we utilized the \textsf{MultiNest} algorithm to scan the theory's parameter space comprehensively. We adopted the leptonic anomalies to guide the scans and included the restrictions from the LHC Higgs data, the DM relic density, the DM direct detection by the XENON-1T experiments, the $B$-physics measurements, and the vacuum stability. We also scrutinized the samples acquired in the scans with the restrictions from the LHC search for SUSY and the latest LZ experiment.
We obtained the conclusions mainly concerning the DM annihilation mechanisms to achieve the measured DM density and the substantial restrictions of the LHC search for SUSY and the DM direct detections on the parameter space.
Most of these conclusions are the same as those in Ref.~\cite{Cao:2022htd}, which similarly studied the muon \texorpdfstring{$g-2$}{} anomaly in the $\mathbb{Z}_3$-NMSSM. However, we still had new observations, summarized as follows:
\begin{itemize}
\item $a_e^{\rm SUSY}$ was mainly correlated with $a_\mu^{\rm SUSY}$ by the formula $a_e^{\rm SUSY}/m_e^2 \simeq a_\mu^{\rm SUSY}/m_\mu^2$ because the two quantities depended similarly on the parameters $M_1$, $M_2$, and $\mu$, and meanwhile, the electron-type and muon-type sleptons were comparable in mass. As a result, $a_e^{\rm SUSY}$ was typically around $5 \times 10^{-14}$ when $a_\mu^{\rm SUSY} \simeq 2.5 \times 10^{-9}$. In addition,  significant violations of this correlation might occur only in rare cases such as $m_{\tilde{\mu}_L} \gg M_2, \mu, m_{\tilde{e}_L}$, where $a_e^{\rm SUSY}/m_e^2$ could be much larger than $a_\mu^{\rm SUSY}/m_\mu^2$.

\item The DM direct detection and LHC experiments were more crucial in determining the results of this research than those in Ref.~\cite{Cao:2022htd}. This conclusion was reflected not only by the percentages of the samples excluded by the restrictions, which were presented in Tables~\ref{Table4} and~\ref{Table5}, but also by the maximum reach of $a_e^{\rm SUSY}$. Concretely, $a_e^{\rm SUSY}$ might be around $3 \times 10^{-13}$ in the optimum cases if one used the XENON-1T experiment to limit the SUSY parameter space. This prediction, however, was reduced to $1.5 \times 10^{-13}$ after implementing the LZ restrictions and $1.0 \times 10^{-13}$ when further considering the LHC restrictions. The underlying reason for this phenomenon is that the electron \texorpdfstring{$g-2$}{} anomaly prefers lighter electroweakinos than the muon \texorpdfstring{$g-2$}{} anomaly. It also requires the presence of light electron-type sleptons. The experiments have tightly limited these situations.

\item On the premise of predicting $a_e^{\rm SUSY} > 0 $ and explaining the muon \texorpdfstring{$g-2$}{} anomaly at the $2 \sigma$ level, the LHC restrictions have set lower bounds on the sparticle mass spectra, e.g., $ m_{\tilde{\chi}_1^0} \gtrsim 220~{\rm GeV}$ and $m_{\tilde{\chi}_2^0}, m_{\tilde{\chi}_1^\pm} \gtrsim 240~{\rm GeV}$, where $\tilde{\chi}_2^0$ and $\tilde{\chi}_1^\pm$ are wino-dominated. Remarkably, although these lower bounds depended on the obtained parameter points or, more basically, their posterior distribution,
    our results coincided with those from the experimental analyses in Fig.~16 of Ref.~\cite{ATLAS:2021moa}, which concluded no LHC restrictions on winos in the $\tilde{B}-\tilde{W}$ co-annihilation case if $m_{\tilde{\chi}_1^0} \gtrsim 220~{\rm GeV}$. The reason was that we accumulated lots of samples in the low $|m_{\tilde{\chi}_1^0}|$ region in the right panel of Fig.~\ref{fig1} and thus explored numerous
    situations of SUSY.

    These bounds can be understood intuitively as follows: if $\tilde{\chi}_1^0$ is lighter, more missing transverse energy will be emitted in the sparticle production processes at the LHC, which can improve the sensitivities of the experimental analyses; while if the sparticles other than $\tilde{\chi}_1^0$ are lighter, they will be more copiously produced at the LHC to increase the events containing multiple leptons.

\item Since the relic density was highly correlated with the DM-nucleon scattering cross-sections in the $Z$- and $h$-funnel regions, one needed great fine tunings of relevant SUSY parameters to yield the measured density without conflicting with the restrictions of the DM direct detection experiments. This problem became more and more severe with the improvement of the sensitivities of the direct detection experiments. As a result, these annihilation scenarios were usually neglected in a less elaborated scan when considering the LZ restrictions.

\end{itemize}

\section*{Acknowledgement}
This work is supported by the National Natural Science Foundation of China (NNSFC) under grant No. 12075076.

\bibliographystyle{CitationStyle}
\bibliography{g-2}

\end{document}